\documentclass[11pt, twoside, a4paper]{article}
\usepackage[margin = 1in]{geometry}
\newcommand{\SC}[1]{ \text{Social Cost}(#1)}
\newcommand{\SW}[1]{ \text{Social Welfare}(#1)}

\usepackage[pdftex]{graphicx}

\usepackage{amsmath, amssymb, amsfonts}
\usepackage{amsthm}
\usepackage{hyperref}       % hyperlinks
\usepackage[utf8]{inputenc} % allow utf-8 input
\usepackage[T1]{fontenc}    % use 8-bit T1 fonts
\usepackage{url}            % simple URL typesetting
\usepackage{booktabs}       % professional-quality tables
\usepackage{amsfonts}       % blackboard math symbols
\usepackage{nicefrac}       % compact symbols for 1/2, etc.
\usepackage{microtype}      % microtypography
\usepackage{comment}

\newcommand{\norm}[2][]{\ensuremath{\left\Vert #2 \right\Vert_{#1}}}

\makeatletter
\newcommand*{\defeq}{\mathrel{\rlap{%
                     \raisebox{0.3ex}{$\m@th\cdot$}}%
                     \raisebox{-0.3ex}{$\m@th\cdot$}}%
                    =}
\newcommand*{\eqdef}{=
  \mathrel{\rlap{%
      \raisebox{0.3ex}{$\m@th\cdot$}}%
    \raisebox{-0.3ex}{$\m@th\cdot$}}%
}
\makeatother

\newcommand{\etal}[0]{\emph{et al. }}

\renewcommand{\vec}[1]{\mathbf{#1}}

\newtheorem{theorem}{Theorem}[section]

\newtheorem{claim}[theorem]{Claim}

\newtheorem{lemma}[theorem]{Lemma}

\newtheorem{corollary}[theorem]{Corollary}
\newtheorem{remark}[theorem]{Remark}

\newtheorem{definition}[theorem]{Definition}

\usepackage{pstricks, sgamevar, egameps}
\usepackage{subfigure, subfloat, graphicx} %epsfig,

\begin{document}

% Page heads
%\markboth{G. Zhou et al.}{A Multifrequency MAC Specially Designed for WSN Applications}

% Title portion
\title{Average Case Performance of Replicator Dynamics in Potential Games via Computing Regions
of Attraction}
\author{Ioannis Panageas \\Georgia Institute of Technology\\ ioannis@gatech.edu
\and Georgios Piliouras \\ Singapore University of Technology \& Design  \\georgios@sutd.edu.sg}
\date{}

\maketitle
\thispagestyle{empty}

\begin{abstract}
What does it mean to fully understand
the behavior of a network of adaptive agents?
 The golden standard typically is the behavior of learning dynamics in potential games, where many
evolutionary dynamics, e.g., replicator dynamics, are known to converge to sets of equilibria. Even in such classic settings
many questions remain unanswered. We examine issues such as:
%\begin{comment}
\begin{itemize}
\item{\textit{Point-wise convergence:}} Does the system always equilibrate, even in the presence of continuums of equilibria?
\item{\textit{Computing
 regions of attraction:}} Given point-wise convergence can we compute %the volume of
 the region of asymptotic stability of each equilibrium (e.g., estimate its volume, geometry)?
  \item{\textit{System invariants:}}
  Invariant functions
  remain constant along every system trajectory.
This notion is orthogonal to the game theoretic concept of a potential function, which always strictly increases/decreases along system
  trajectories.  Do dynamics in potential games exhibit invariant functions? If so, how many? How do these functions look like?
  \end{itemize}
Based on these geometric characterizations,  we propose a novel
quantitative framework for analyzing the efficiency of potential games with many equilibria.
The predictions of different equilibria are weighted by their probability to
arise under evolutionary dynamics
 given uniformly random initial conditions.
 This average case analysis is shown to offer novel insights
 in classic game theoretic challenges, including quantifying the risk dominance
 in stag-hunt games and
  allowing for more nuanced performance analysis
  in networked coordination and congestion games with  large
  gaps between
   price of stability and price of anarchy.
%\end{comment}
\end{abstract}

%ACM is moving forward with the 2012 Classification system: http://dl.acm.org/ccs_flat.cfm. Please generate the CCSXML tex code through the online interactive system and insert the code below.

%\maketitle

\section{Introduction}

The study of game dynamics is a basic staple of game theory
%The history of the field traces back to the early work of Brown and Robinson \cite{Brown1951,Robinson1951} on learning dynamics in zero-sum games,
%which itself came shortly on the heels of von Neumann's foundationary investigation of equilibrium computation.
%Since then vast amount of work has accumulated on the subject
with several books dedicated exclusively to it
 \cite{Hofbauer98,Fudenberg98,Young,Cesa06,Sandholm10}. %,Cesa06,Fudenberg98,
%
\begin{comment}
For example, in the case of zero-sum games, it is well known that a large number of dynamics (e.g., no regret dynamics) converge in a weak sense to the set of Nash equilibria. Specifically,  the time average of play converges to the set of Nash equilibria. However, this clearly leaves many possibilities open about the actual day-to-day system behavior. Is the system periodic? Can it exhibit more complex behavior? The answers to these questions about the actual system trajectories tend to be rather complicated  \cite{akin,Sato02042002}. %Soda14,PiliourasAAMAS2014}.
In order to avoid such disequilibrium issues,
\end{comment}
%
%A key question when it comes to the study of any system is what constitutes a sufficient understanding of its behavior.
%In the case of learning in games the
Historically, the golden standard for classifying the behavior of learning dynamics in games
  has been to establish convergence  to equilibria. Thus,  it is hardly surprising that a significant part of the work on learning in games focuses on potential games (and slight generalizations thereof) where many dynamics (e.g., replicator, smooth fictitious play) are known to converge to equilibrium sets. The structure of the convergence proofs is essentially universal across different learning dynamics and boils down to identifying a Lyapunov/potential function that strictly decreases along any nontrivial trajectory. In potential games, as their name suggests, this function is part of the description of the game and precisely guides self-interested dynamics towards critical points of these functions that correspond to equilibria of the learning process.

Potential games are also isomorphic to congestion games \cite{potgames}. Congestion games have been instrumental in the study of efficiency issues in games. They are amongst the most extensively studied class of games from the perspective of price of anarchy and price of stability with many tight characterization results for different subclasses of games (e.g., linear congestion games \cite{Roughgarden09}, symmetric load balancing \cite{Nisan:2007:AGT:1296179} and references therein).

Given this extensive treatment of such a classic class of games it would seem, at a first glance, that our understanding of these systems is more or less complete. We show that this is far from the case. We focus on simple systems where replicator dynamic, arguably one of the most well studied game dynamics, is applied to linear congestion games and (network) coordination games.  We resolve a number of basic open questions in  the following results:
%along with new benchmarks and yardsticks for a much tighter understanding of learning in games.
%Our hope is that these approaches
% may inspire different vantage points in the study of learning in games and
 %may inspire similarly detailed questions in
%can be pursued in wider classes of learning in games and hopefully more generally in different settings of networked computation (such as in the study of biological evolution, see related work section).
%
%\textbf{Our results:}
%Our results can be organized around four key axes: point-wise convergence to equilibrium, regions of attraction, average case performance analysis, and  system invariants. We examine each of them next:
 %Each of them can be understood as a more nuanced study of well established themes in the area. % of learning in games.

\textbf{A) Point-wise convergence to equilibrium:} In the case of  linear congestion games and (network) coordination games
%We focus on  potential games which are known to be isomorphic to congestion games \cite{potgames}.
we prove convergence to equilibrium instead of equilibrium sets.
%We start by elucidating the difference between these two statements.
%Although they are virtually identical from a linguistic perspective, they are quite distinct from a formal, topological perspective.
 Convergence to equilibrium sets implies that the distance of system trajectories from the sets of equilibria converges to zero. On the other hand, convergence to equilibrium, also referred to as  point-wise convergence, implies that every  system trajectory has a unique limit point, which is an equilibrium. In games with continuums of equilibria, (e.g., $N$ balls $N$ bins games\footnote{These are symmetric load balancing games with $n$ agents and $n$ machines where the cost function of each machine is the identity function.} with $N\geq4$), the first statement is more inclusive that the second. In fact, system equilibration is not implied by set-wise convergence, and the limit set of a trajectory may have complex topology (e.g., the limit of social welfare may not be well defined).
%\textit{We establish point-wise convergence for replicator dynamics
%in general linear congestion games and arbitrary networks of coordination games.}
Despite numerous positive convergence results in classes of congestion games (\cite{Fotakis08,Berenbrink12,Even-Dar:2005:FCS:1070432.1070541,Berenbrink:2007:DSL:1350525.1350533,Ackermann:2009:CID:1582716.1582732}),
  this is the first to our knowledge result about deterministic point-wise convergence for any concurrent dynamic.
%To our knowledge this is the first result of deterministic point-wise convergence to equilibria in linear congestion games for \textit{any} concurrent dynamic.
 This argument is based on combining  global Lyapunov functions arguments with  local information theoretic  Lyapunov functions around each equilibrium.

\textbf{B) Global stability analysis:}
Although the point-wise convergence result is interesting in itself, it critically enables all other results in the paper. Specifically, we establish that modulo point-wise convergence,  all but a zero measure set of initial conditions converge to equilibrium points which are  (linearly) stable ($i.e.$, their Jacobian has no eigenvalue with positive real part). This is a technical result that combines game theoretic arguments with tools from dynamical systems (Center-Stable Manifold theorem) and analysis (Lindel\H{o}f's lemma). \cite{Kleinberg09multiplicativeupdates} has established that all such equilibria must satisfy a refined game theoretic property and known as weakly stability.  A Nash equilibrium is weakly stable if given any two randomizing agents, fixing one of the agents to choosing one of his strategies with probability one, leaves the other agent indifferent between the strategies in his support. This condition is easy to work with (does not require computing eigenvalues) and sometimes (along with the global stability result) already suffices to make a unique prediction about the resulting system performance.

\textbf{C) Invariant functions:}
Sometimes a game may have multiple (weakly) stable equilibria. In this case we would like to be able to predict which one will arise given  a specific (or maybe a randomly chosen) initial condition. Systems invariants allows us to do exactly that.
 A system invariant is a function defined over the system state space such that it remains constant along every system trajectory.
Establishing invariant properties of replicator dynamics in generalized zero-sum games has helped prove interesting topological properties of the system trajectories such as (near) cycles \cite{PiliourasAAMAS2014, Soda14,Papadimitriou:2016:NEC:2840728.2840757}. In the case of bipartite coordination games with fully mixed Nash equilibria, we can establish similar invariant functions. Specifically,
 the difference between the sum of the Kullback-Leibler (K-L) divergences of the evolving mixed strategies of the agents on the left partition from their fully mixed Nash equilibrium strategy and the respective term for the agents in the right partition remains constant along any trajectory. In the special case of star graphs, we show how to produce $n$ such invariants where $n$ is the degree of the star. This allows for creating efficient oracles for predicting to which Nash equilibrium the system converges provably for \text{any} initial condition without simulating explicitly the system trajectory.

 \textbf{Applications:} The tools that we have developed allow for novel insights in classic and well studied class of games. We group  our results into two clusters, average case performance analysis and estimating risk dominance/regions of attraction.

 \textit{Average Case Performance:} We propose a novel
quantitative framework for analyzing the efficiency of potential games with many equilibria.
 %point of view
Informally, we define the expected system performance as the weighted average of the social costs
of all equilibria where the weight of each equilibrium is proportional to the volume (or more generally measure) of its region of attraction.

The main idea is as follows: The agents start participating in the game having some prior beliefs about which are the best actions for them.
We will typically assume that the initial beliefs are chosen according to a uniform prior given that we want to assume no knowledge about the agents' internal beliefs\footnote{Our techniques extend to arbitrarily correlated beliefs, any prior over initial mixed strategies.}. Given this initial condition the agents start interacting through the game and update their beliefs (\textit{i.e.}, their randomized strategies)  up until they reach equilibrium. At this point the measure of the region of attraction of an equilibrium captures exactly the likelihood that we will converge to that state. So the average case performance computes, as its names suggests, what will be the resulting system performance on average. As is typical in algorithmic game theory, we can normalize this quantity by dividing with the performance of the optimal state. We define this ratio as the average price of anarchy. In our convergent systems it always lies between the price of stability and the price of anarchy.
We analyze the average price of anarchy in a number of settings which include, $N$ balls $N$ bins games, symmetric linear load balancing games (with agents of equal weights)\footnote{We focus mostly on the makespan as a measure of social cost.}, parametric versions of coordination games as well as star network extensions of them. These are games with large gaps between the price of stability and price of anarchy and replicator  is shown to be able to zero in on the good equilibria with high enough probability so that the average price of anarchy is always a small constant. This measure of performance could help explain why some games are  easy in practice, despite having large price of anarchy. We aggregate these results below:

\begin{table}[h!]
  \centering
  \resizebox{\textwidth}{!}{
    \begin{tabular}{ | l | c | c | c| c | c | }
    \hline
   &Average PoA & Techniques & PoS &  Pure PoA & PoA  \\ \hline
  $N$ balls $N$ bins game   & $\textbf{1}$              &\textbf{A \& B}                       & $1$ &$1$ & $\Theta (\log n/\log\log n)$  \\ \hline
 Symmetric Load Balancing    & $\textbf{[1, 1.5]}$  &\textbf{A \& B}                         & $1$ &$1$ & $\Omega(\log n/\log\log n)$  \\ \hline
  $w$-Coordination Game  & $\textbf{[1.15, 1.21]}$  &\textbf{A \& B \& C}                       &$1$ & $\Theta(w)$ & $\Theta(w)$ \\ \hline
  $N$-Star $w$-Coordination Game &  $\textbf{[1.15, 1.42]}$ &\textbf{A \& B \& C}        &  $1$ & $\Theta(w)$ & $\Theta(w)$  \\
    \hline
    \end{tabular}}

    \end{table}

\begin{comment}
In order to get an intuitive understanding of these issues, let's examine a variant of  Zeno's paradox of the tortoise and Achilles. In a twist of the typical narrative, now at the end of the $t$-phase of the run, instead of the tortoise gaining a $1/2^t$ advantage over Achilles, our faster tortoise has a $1/t$  length advantage over Achilles. In this case, the series diverges and Achilles actually does not reach the tortoise (i.e., his trajectory up to reaching the tortoise is infinitely long). Similarly, we may have trajectories that converge to an equilibrium set, however the %rate of
speed with which they move converges to zero slowly. It is still possible for such trajectories to have infinite length and span complex patterns in the vicinity of a continuum of an equilibrium set. The system behavior may be rather complex, despite the seemingly reassuring convergence to equilibrium sets statement. In fact, such a statement is borderline inaccurate since %at a linguistic level
it invites the false presumption that the system actually converges, when in fact it may well not!
\end{comment}

%Although these appears as a minor, almost syntactic point

 % around each equilibrium and not on the typical global potential functions used in establishing the standard  set-wise convergence to equilibria.
%This  result, which is of independent interest, allows us to define properly the notion of average case system performance  where the social welfare of each equilibrium is weighted by the size of its basin.

 \textit{Risk dominance/Regions of attraction:}
 Risk dominance is an equilibrium refinement process that centers around uncertainty about opponent behavior.
A Nash equilibrium is considered risk dominant if it has the largest basin of attraction\footnote{Although risk dominance \cite{Selten88}  was originally introduced as a hypothetical model of the method by which perfectly rational players select their actions,
 it may also be interpreted \cite{Myerson91}  as the result of evolutionary processes.}.
The benchmark example is the Stag Hunt game, shown in figure \ref{fig:game1}.
In such symmetric 2x2 coordination games a strategy is risk dominant if it is a best response to the uniformly random strategy of the opponent.
  We show that the likelihood  %size of attraction
 of the risk dominant equilibrium of the Stag Hunt game is
 $\frac{1}{27} (9 + 2 \sqrt{3} \pi) \approx 0.7364$ (instead of merely knowing that it is at least $1/2$, see figure \ref{fig:game13}).
 The size of the region of attraction of the risk dominated equilibrium is $0.2636$, whereas the mixed equilibrium
 has region of attraction of zero measure.
Moving to networks of coordination games, we show how to construct an oracle that predicts the limit behavior of an arbitrary initial condition, in the case of coordination games played over a star network with $N$ agents. This is the most economic class of games that exhibits two characteristics that intuitively seem to pose intractable obstacles to the quantitative analysis of nonlinear systems: i) they have   (arbitrarily many) free variables, ii) they exhibit a continuum of equilibria.

\section{Related Work}

\textbf{Set-wise convergence in congestion/potential games:}
%In \cite{rosenthal73} Rosenthal showed that every congestion game has pure Nash equilibria, and that better-response dynamics converge to them. In these dynamics,
 %in every round, exactly one agent deviates to a better strategy. If two or more agents move at the same time then convergence is not guaranteed.
 A number of positive convergence results have been established for concurrent dynamics \cite{Fotakis08,Berenbrink12,Even-Dar:2005:FCS:1070432.1070541,Berenbrink:2007:DSL:1350525.1350533,Ackermann:2009:CID:1582716.1582732,Kleinberg09multiplicativeupdates}, however, they usually depend on strong assumptions about network structure (e.g., load balancing games) and/or symmetry of available strategies and/or are probabilistic in nature and/or establish convergence to approximate equilibria. On the contrary our convergence results are deterministic, hold for any network structure and are \textit{point-wise}.  %, which is the strongest topological convergence statement that one can hope to prove in this class of games.

\textbf{Learning as a refinement/prediction mechanism in game theory:}
\begin{comment}
A Nash equilibrium  is evolutionarily stable if  once it is fixed in a population, natural selection alone is sufficient to prevent alternative (mutant) strategies from invading successfully \cite{smith}.
%Such fixed points are referred to as evolutionary stable equilibria or evolutionary stable strategies \cite{smith}.
%volutionary stable equilibia (or evolutionary stable strategies)
A related concept is that of an evolutionary stable state \cite{smith2}. This is a definition that arises in mathematical biology and explicitly in the study of single population replicator dynamics.  A population is said to be in an evolutionarily stable state if its genetic composition is restored by selection after a disturbance, provided the disturbance is not too large.
A stochastically stable equilibrium \cite{foster} is a further refinement of the evolutionarily stable states. An evolutionary stable state is also stochastically stable if under vanishing noise the probability that the population is in the vicinity of the state  does not go to zero.
\end{comment}
Price of anarchy-like bounds in potential games using equilibrium stability refinements (e.g., stochastically stable states)  have been explored before
\cite{Chung:2008:PSA:1483688.1483722,SaberiPoA,Ackermann:2009:CID:1582716.1582732}.  Our
approach and techniques are more expansive in scope, since they also allow for computing the actual likelihoods of each equilibrium as well as the topology of the regions of attractions of different equilibria.
%
%\smallskip
%
%\textbf{Replicator dynamics and system performance:}

We build upon positive performance results  %that show favorable performance guarantees
 for replicator dynamics (and discrete-time variants). % in settings of interest.
The key reference is \cite{Kleinberg09multiplicativeupdates}, where many key ideas including the fact that replicator dynamics can significantly outperform
worst case equilibria were introduced.
This stability analysis can be generalized to deterministic variants of replicator~\cite{ITCS15MPP}.
Replicator  can outperform even best case equilibria by converging to cycles~\cite{paperics11,sigecom11}.
 %Finally, the use of information theoretic arguments as (weak) Lyapunov functions has been explored before (\textit{e.g.}, \cite{paperics11,kleinberg2011load}).
% Head 2

In independent parallel work \cite{Zhang2014} examine equilibrium selection issues in $2\times2$ coordination games for replicator dynamics, however, their techniques do not scale to larger games.
Analyzing the regions of attraction for (variants of) replicator dynamics in (time-evolving) games raises interesting computational questions relevant to mathematical biology \cite{2014arXiv1411.6322M,2015arXiv151101409M}.

\cite{Papadimitriou:2016:NEC:2840728.2840757} show how to use elements from the theory of topology of dynamical systems such as chain recurrent sets to analyze learning dynamics in games. This solution concept generalizes the notion of Nash equilibrium and captures the actual limit behavior of game dynamics. % can be significantly larger than the set of all Nash equilibria.
 Combining the ideas of regions of attraction with chain recurrent sets opens up interesting directions for average case analysis of learning dynamics in non-potential games.

 \textbf{Regions of Attraction \& Gradient Dynamics in Non-Convex Optimization:}
 In recent work, \cite{DBLP:journals/corr/LeeSJR16,DBLP:journals/corr/PanageasP16} have shown how to combine tools from dynamical systems theory to understand the behavior of one of the most classic optimization heuristics, deterministic discrete-time fixed step-size gradient dynamics in general non-convex fitness landscapes. Specifically, it is argued that saddle points (non-local minima fixed points) have regions of attraction of zero measure and hence gradient dynamics typically converge to local minima.

% Say that convergence to a set is different from pointwise convergence

\subsection*{Organization of the paper}
The rest of the paper is organized as follows. In Section \ref{sec:definitions} we provide definitions in regards to dynamical systems, congestion and network coordination games and the average price of anarchy. In Section \ref{sec:tools} we establish point-wise convergence of replicator dynamics for congestion and network coordination games and we develop the mathematical machinery necessary for approximating the average price of anarchy. In Section \ref{sec:average} we present our average price of anarchy results. All the missing proofs can be found in the appendix.
\section{Definitions and Basic Tools}\label{sec:definitions}
\subsection*{Notation}
 We use boldface letters, e.g. $\vec{x}$, to denote vectors and denote a vector's $i^{th}$ coordinate by $x_i$. We use $\vec{x}_{-i}$ to denote $\vec{x}$ after removing coordinate $i$-th. For a function $f,$ we denote by $f^n$  the composition of $f$ with itself $n$ times, namely $\underbrace{f \circ f \circ \cdots\circ f}_{n \textrm{ times}}$. We use $\mathcal{J}[\vec{x}]$ to denote the Jacobian matrix (of some function clear from the context) at the point $\vec{x}.$
\subsection{Dynamical Systems}\label{sec:systems}
Let $f:\mathcal{S} \to \mathbb{R}^n$ be continuously differentiable with $\mathcal{S} \subset \mathbb{R}^n$, $\mathcal{S}$ an open set. We examine \emph{continuous (time) dynamical systems} of the form
\begin{equation}\label{eq:continuous}
\frac{d \vec{x}}{dt} = f(\vec{x}).%,\; \vec{x}_{t+1} = g(\vec{x}_{t})
\end{equation}
Since $f$ is continuously differentiable, the ordinary differential equation (ode (\ref{eq:continuous})) along with the initial condition $\vec{x}(0) = \vec{x}_0 \in \mathcal{S}$ has a unique solution for $t \in \mathcal{I}(\vec{x}_0)$ (some time interval) and we can present it by $\phi(t,\vec{x}_0)$, called the \emph{flow} of the system. $  \phi_{t}(\vec{x}_0) \defeq \phi(t,\vec{x}_0)$ corresponds to a function of time which captures the \emph{trajectory} of the system with $\vec{x}_0$ the given starting point. It is continuously differentiable, its inverse exists (denoted by $\phi_{-t}(\vec{x}_0)$) and is also continuously differentiable (called \emph{diffeomorphism}) in the so called maximal interval of existence $\mathcal{I}$. It is also true that $\phi_{t}\circ \phi_{s} =\phi_{t+s}$ for $t,s,t+s \in \mathcal{I}$ and therefore $\phi_{k} = \phi_1^k$ for $k \in \mathbb{N}$ (composition of $\phi_1$ $k$ times as long as $1,k \in \mathcal{I}$). $\vec{x}_0 \in \mathcal{S}$ is called an \emph{equilibrium} if $f(\vec{x}_0)=\vec{0}$. In that case holds $\phi_t(\vec{x}_0) = \vec{x}_0$ for all $t\in \mathcal{I}$, i.e., $\vec{x}_0$ is a \emph{fixed point} of the function $\phi_{t}(\vec{x})$ for all $t\in \mathcal{I}$. We call $\vec{x}_0$ \emph{linearly stable} if the eigenvalues of the Jacobian $\mathcal{J}[\vec{x}_0]$ of $f$ (at the fixed point $\vec{x}_0$) have non-positive real part.

%\textbf{Remark.} 
 If $f$ is globally Lipschitz then the flow is defined for all $t \in \mathbb{R}$, i.e., $\mathcal{I} = \mathbb{R}$. One way to enforce the dynamical system to have a well-defined flow for all $t \in \mathbb{R}$ is to renormalize the vector field by $\norm[]{f(\vec{x})}+1$, i.e., the resulting dynamical system will be $\frac{d \vec{x}}{dt} = \frac{f(\vec{x})}{\norm[]{f(\vec{x})}+1}$, because the function becomes globally $1$-Lipschitz.
The two dynamical systems (before and after renormalization) are topologically equivalent (\cite{perko}, p. 184). Formally this means that there exists a homeomorphism $H$ which maps trajectories of (\ref{eq:continuous}) onto trajectories of the renormalized flow and preserves the direction of time. In words it means that the two systems have the same behavior/geometry (same fixed points, convergence properties, phase portrait).
%As far as discrete time dynamical systems are concerned, $\vec{x}_0 \in \mathcal{S}$ is called a \emph{fixed point} or \emph{equilibrium} if %$g(\vec{x}_0) = \vec{x}_0$ and a \emph{trajectory} with starting point $\vec{x}_0$ is the \emph{sequence} $(g^t(\vec{x}_0))_{t %\in\mathbb{N}}.$ Also $\vec{x}_0$ is called \emph{linearly stable} if the eigenvalues of the Jacobian of $g$ $J[\vec{x}_0]$ (at the fixed %point $\vec{x}_0$) have absolute value at most 1.

A Lyapunov (or potential) function $V: \mathcal{S} \to \mathbb{R}$ is a function that strictly decreases along every non-trivial trajectory of the dynamical system. Formally, for continuous time dynamical systems it holds that $\frac{dV}{dt}\leq 0$ with equality only when $f(\vec{x})=\vec{0}$.
%For discrete time dynamical systems, it is true that $V(\vec{x}) \geq V(g(\vec{x}))$ with equality only at fixed points.
For more information on dynamical systems see \cite{perko}.

\subsection{Average Performance of a system}\label{sec:measure}
Let $\mu$ be the Lebesgue measure on $\mathbb{R}^n$ and assume that $\mu(\mathcal{S})>0$. Given a dynamical system (continuous time) we assume that $\lim_{t\to \infty}\phi_t(\vec{x})$
%, \lim_{t\to \infty}g^{t}(\vec{x})$
exists for all $\vec{x} \in \mathcal{S}$ (the limit is called a \emph{limit point}); the system \emph{converges} \emph{point-wise} for all initial conditions. In this case, continuity implies that every trajectory converges to some equilibrium (fixed point) of the dynamics.\footnote{If $\lim_{t \to \infty} h^t(\vec{x})=\vec{y}$ and $h$ continuous then $h(\vec{y})=\vec{y}.$ Set $h \defeq \phi_1$.}

We would like to understand the average (long-term) behavior of the convergent system (e.g., if the initial condition is chosen uniformly at random from $\mathcal{S}$). Intuitively, since the system converges to fixed points, we would like each fixed point to be assigned weight proportional to its \emph{region of attraction}. We define the region of attraction of a fixed point $\vec{x}_0$ by $R_{\vec{x}_0} = \{\vec{x} \in \mathcal{S}: \lim_{t \to \infty}\phi_{t}(\vec{x}) = \vec{x}_0\},$ namely the set of starting points so that the dynamic converges to $\vec{x}_0$.
Let $\psi(\vec{x}) = \lim_{t\to \infty}\phi_t(\vec{x})$, i.e., $\psi$ maps each starting point $\vec{x}$ to the limit of the $\phi_t(\vec{x})$. It turns out that $\psi$ is measurable (see Lemma \ref{lem:measurable}) and we can define the average (long-term) performance of the system under some (utilitly/cost) function $u$. Let $u: \mathcal{S} \to \mathbb{R}$ be continuous then the \emph{average (case) performance} of a system is defined as \begin{equation}\label{eq:average}
\textrm{acp}_u \defeq \frac{\int_{\mathcal{S}} u\circ \psi d\mu}{\mu(\mathcal{S})} = \mathbb{E}_{\vec{x} \sim U(\mathcal{S}) }[u(\psi(\vec{x}))],
 \end{equation}
 where $U(\mathcal{S})$ is the uniform distribution on $\mathcal{S}$. $u$ quantifies the quality of the points $\vec{x} \in \mathcal{S}$ (e.g., social welfare in games). Observe that if $m \defeq \min_{\vec{x} \in \textrm{FP}} u(\vec{x}), M \defeq \max_{\vec{x} \in \textrm{FP}} u(\vec{x})$ where FP denotes the set of fixed points\footnote{The set of fixed points in $\mathcal{S}$ is closed.} then $m \leq \textrm{acp}_u \leq M$ (a). We believe that computing/approximating the average case performance is an important step towards understanding the actual behavior of a system.

%Let $R_{\mathcal{S}'} = \{\vec{x} \in \mathcal{S}:=\vec{y} \textrm{ and } \vec{y}\in \mathcal{S}'\}$ i.e $R_{\mathcal{S}'}$ is the region of %attraction of set $S'$. We define the following probability measure $\nu$, so that $\nu(S') = \frac{\mu(R_{S'})}{\mu(S)}$. Measure $\nu$ is %well-defined because $R_{S'}$ is a (Lebesgue) measurable set for all $S'\subseteq S$ ($\phi_{1}$ are continuous) and satisfies the properties %of a measure (see Lemma \ref{lem:measure}). Analogously, we can define the $\nu$ for discrete time dynamical systems that converge pointwise. %Think of $\nu$ as a probability distribution over fixed points, where fixed points with more measure are more ``likely" (under uniformly %random initial starting point) to be reached by the dynamical system in the long-term. In particular, if $S' = \{\vec{x}_0\}$ where %$\vec{x}_0$ is a fixed point of the dynamics, then by $\nu$'s definition we have that $\nu(\{\vec{x}_0\})$ is proportional to the  region of %attraction of $\vec{x}_0$ (sometimes we denote $\nu(\{\vec{x}_0\})$ by $\nu(\vec{x}_0)$).

%Knowing $\nu$, we can describe the long-term behavior of the system, without knowing the equations of the dynamics explicitly!
To see the connection with game theory, think of $\mathcal{S}$ as the set of mixed (randomized) strategies, a fixed point with region of attraction of positive measure as a Nash equilibrium and $u$ as the social cost/welfare. In this case, integral (\ref{eq:average}) becomes a weighted average among the social cost/welfare of the Nash equilibria. The average case performance is sandwiched between the values (of the social cost/welfare) at the worst, best Nash equilibrium.

We use (continuous time) \emph{replicator dynamics} on \emph{congestion and network coordination games} as our benchmark. In this case, the set of Nash equilibria is a subset of the set of fixed points. Nevertheless, we can show that the dynamics converge point-wise and finally that Nash equilibria are the only fixed points whose region of attraction may be of positive Lebesgue measure. Later in this section we define the notion of \emph{average price of anarchy} which is essentially a scaled version of average performance, defined particularly for games.
\begin{remark}[Generalizations of average case performance] The definition of average case performance can be used for any point-wise convergent discrete time dynamical systems (function $\psi(\vec{x})$ will be equal to $\lim_{k \to \infty}g^{k}(\vec{x})$ where $g$ is the rule of the discrete dynamics). Also, different measures of efficiency can be defined where the initial condition follows some distribution other than the uniform. Generally, the distribution over initial conditions, the notion of (social) utility/cost, and the dynamic can all be treated as parameters of this performance measure.
\end{remark}
%The dynamics we focus on is called Replicator Dynamics for the class of Linear Congestion Games. In this case, the stable fixed points of the %dynamics are Nash Equilibria and $\nu$ can be seen as a probability distribution over Nash equilibria. This is true because we will show that %$\mu(S')=0$ for any $S'$ so that $S'$ does not

%As we can see later, $\nu$
%if $\vec{x}$ is a fixed point and $\nu(\{\vec{x}\})=0$ otherwise. Finally $\nu(S') = \nu_{\vec{x} \in S'}(\{\vec{x}\})$
% . It is not hard to show that in case of convergent systems, $R_{\vec{x}}$ is measurable, i.e $\mu(R_{\vec{x}})$ is well defined.
%Say why this is important for average perfomance equlibrium selection and understanding the behavior of a system.
%Under the case that Nash equilibria == fixed points with positive measure then this captures the average performance of a system. Replicator dynamics that we define formally below when applied on congestion and coordination games , we have this property!!
\subsection{Replicator Dynamics on Congestion/Network Coordination Games}\label{sec:prelims2}
\subsubsection*{Congestion Games}\label{sec:congestion}
%Congestion games~\cite{rosenthal73} are non-cooperative games in which the utility of each agent depends only on
%the agent's strategy and the number of other agents that either choose the same strategy, or
%some strategy that  intersects/overlaps it.
A  \emph{congestion game}  is defined by the tuple $(\mathcal{N}; \mathcal{E}; (S_i)_{i \in \mathcal{N}}; (c_e)_{e \in \mathcal{E}})$ where $\mathcal{N}$ is the set of \emph{agents} (with $N = |\mathcal{N}|$),  $\mathcal{E}$
is a set of \emph{resources} (also known as \emph{edges} or \emph{bins} or \emph{facilities}), and
each player $i$ has a set $S_i$ of subsets of $\mathcal{E}$
($S_i \subseteq 2^{\mathcal{E}}$) and $|S_i|\geq 2$.
Each strategy $s_i \in S_i$
is a set of edges (a \emph{path}), and
$c_e$ is a cost (negative utility)
function associated with facility $e$.
We will also use small greek characters like $\gamma, \delta$ to denote different strategies/paths.
For a  strategy profile $\vec{s} = (s_1,s_2,\dots,s_N)$, the cost
of player $i$ is given by $c_i(\vec{s}) = \sum_{e \in s_i} c_e(\ell_e(\vec{s}))$, where $\ell_e(\vec{s})$ is the
number of players using $e$ in $\vec{s}$ (the load of edge $e$).
 In linear congestion games, the latency functions are of the form  $c_e(x)=a_ex+b_e$ where $a_e,b_e\geq 0$. Measures of social cost ($\textrm{sc}(\vec{s})$)
 include the makespan, which is equal to the cost of the most expensive path and
 the sum of the costs of all the agents.
\subsubsection*{Network (Polymatrix) Coordination Games}\label{sec:network}
A coordination (or partnership) game is a two player game where in each strategy outcome both agents receive the same utility. In other words, if we flip the sign of the utility of the first agent then we get a zero-sum game. An $N$-player polymatrix (network) coordination game is defined by an undirected graph $G(V,E)$ with $|V|=N$ vertices and each vertex corresponds to a player. An edge $(i,j)\in E(G)$ corresponds to a coordination game between players $i,j$. We assume that we have the same strategy space $S$ for every edge. Let $A_{ij}$ be the payoff matrix for the game between players $i,j$ and $A_{ij}^{\gamma\delta}$ be the payoff for both (coordination) if $i,j$ choose strategies $\gamma,\delta$ respectively. The set of players will be denoted by $N$ and the set of neighbors of player $i$ will be denoted by $N(i)$. For a  strategy profile $\vec{s} = (s_1,s_2,\dots,s_N)$, the utility
of player $i$ is given by $u_i(\vec{s}) = \sum_{j \in N(i)} A_{ij}^{s_{i}s_{j}}$.
%$\Delta_{i}$ is the simplex for player $i$ (standard $(|S|-1)$-simplex), $\Delta = \times_{i} \Delta_{i}$.
%The probability player $i$ chooses strategy $\gamma$ is denoted by $p_{i\gamma}$. $u_{i\gamma}$ corresponds to the expected utility %of player $i$ given that he chooses strategy $\gamma$ and is equal to $\sum_{j \in N(i)}A_{ij}^{\gamma\delta}p_{j\delta}$. Moreover, %$\hat{u}_{i}$ corresponds to the expected utility, i.e., $\hat{u}_{i} = \sum_{\gamma}u_{i\gamma}p_{i\gamma}$.
%
 The social welfare of a state $\vec{s}$ corresponds to the sum of the utilities  of all the agents $sw(\vec{s}) = \sum_{i \in V} u_{i}(\vec{s})$.

The \emph{price of anarchy} is  defined as:
% the price of anarchy of game $G$ as:
%% FINISH THIS
%
%PNE may not be efficient from a social perspective.
%A standard measure of distributed inefficiency is % due to selfishness is
%the \emph{price of anarchy} (PoA)~\cite{KoutsoupiasP99WorstCE},
%defined as the ratio of the social cost of the worst PNE to the optimum:
%
%
$
\text{PoA} \!=\!
\displaystyle\frac
{\max\nolimits_{\vec{s} \in \text{NE}} \SC{\vec{s}}}
{\min\nolimits_{\vec{s}^{\ast} \in \times_i S_i} \SC{\vec{s}^{\ast}}}
$ for cost functions and similarly
$
\text{PoA} \!=\!
\displaystyle\frac
{\max\nolimits_{\vec{s}^{\ast} \in \times_i S_i} \SW{\vec{s}^{\ast}}}
{\min\nolimits_{\vec{s} \in \text{NE}} \SW{\vec{s}}}
$
for utilities.\footnote{NE denotes the set of Nash equilibria.}

We denote by $\Delta(S_{i}) = \{\vec{p}\geq \vec{0}: \sum_{\gamma}p_{i\gamma}=1\}$ the set of mixed (randomized) strategies of player $i$ and $\Delta = \times_i \Delta(S_i)$ the set of mixed strategies of all players. For congestion games we use $c_{i\gamma}=\mathbb{E}_{s_{-i}\sim \vec{p}_{-i}} c_i(\gamma,\vec{s}_{-i})$ to denote the expected cost of player $i$ given that he chooses strategy $\gamma$ and $\hat{c}_{i} = \sum_{\delta \in S_i}p_{i\delta}c_{i\delta}$ to denote his expected cost. Similarly, for network coordination games we use $u_{i\gamma}=\mathbb{E}_{s_{-i}\sim \vec{p}_{-i}} u_i(\gamma,\vec{s}_{-i})$ to denote the expected utility of player $i$ given that he chooses strategy $\gamma$ and $\hat{u}_{i} = \sum_{\delta \in S_i}p_{i\delta}u_{i\delta}$ to denote his expected utility.
\subsubsection*{Replicator Dynamics}\label{sec:replicator}
Replicator dynamics %~\cite{Taylor1978145,Schuster1983533,Kleinberg09multiplicativeupdates}
is described by the following system of differential equations adjusted to cost games (e.g., congestion games) and utility games (e.g., network coordination games) respectively:

\begin{equation}
\label{eq:system}
\frac{d p_{i\gamma}}{dt}=p_{i\gamma}\big( \hat{c}_{i} - c_{i\gamma} \big), \;\;\frac{d p_{i\gamma}}{dt}=p_{i\gamma}\big(  u_{i\gamma}- \hat{u}_{i} \big)
\end{equation}
%
 %\noindent
for each $i \in \mathcal{N}$, $\gamma \in S_i$.
 Observe that if $\hat{c}_{i} > c_{i\gamma}$ then $\frac{d p_{i\gamma}}{dt}>0$, i.e., $p_{i\gamma}$ is increasing with time, thus player $i$ tends to increase the probability he chooses strategy $\gamma$. Similarly if $\hat{c}_{i} < c_{i\gamma}$ then $\frac{d p_{i\gamma}}{dt}<0$, i.e., $p_{i\gamma}$ is decreasing with time, thus player $i$ tends to decrease the probability he chooses strategy $\gamma.$\footnote{Replicator dynamics describes rational behavior in a sense.} Replicator dynamics capture similarly rational behavior in the case of network coordination games.
\begin{remark}\label{rem:fixednash}
The fixed points of replicator dynamics are exactly the set of randomized strategies such that each agent experiences equal costs across all strategies he chooses with positive probability. This is a generalization of the notion of Nash equilibrium, since Nash equilibria furthermore require that any strategy that is played with zero probability must have expected cost at least as high as those strategies which are played with positive probability. Moreover, due to the uniqueness theorem for solutions of ordinary differential equations, we have that the flow of replicator dynamics is defined for all $t \in \mathbb{R}$ and initial conditions in $\Delta$ \cite{NowakBook}.
\end{remark}
\subsubsection{Definition of average price of anarchy (APoA)}\label{sec:defapoa}
In this section we define the notion of average price of anarchy, following the machinery from Section \ref{sec:measure}. It is natural to set $\mathcal{S}$ to be the product of simplexes $\Delta$, but this does not suffice since $\Delta$ has measure zero in $\mathbb{R}^M$, where $M \defeq \sum_{i}|S_{i}|$. The reason is that the probabilities sum up to one for each player. To circumvent this issue (since from Section \ref{sec:measure} we need $\mu(\mathcal{S})>0$), we consider a natural projection $g$ of the points $\vec{p} \in \Delta$ to $\mathbb{R}^{M-N}$ by excluding a specific but arbitrarily chosen\footnote{Choose an arbitrary ordering of the strategies of each agent and then exclude the last strategy.} variable for each player. We denote $g(\Delta)$ the ``projected" product of simplexes and the projection of any point $\vec{p} \in \Delta$ by $g(\vec{p})$ (for example $(p_{1,a},p_{1,b},p_{1,c},p_{2,a'},p_{2,b'}) \to_{g} (p_{1,a},p_{1,b},p_{2,a'})$ where $p_{1,a}+p_{1,b}+p_{1,c}=1$ and $p_{2,a'}+p_{2,b'}=1$)). Given a dynamical system\footnote{We assume that this system describes the evolution of the mixed strategies of rational agents in some game.} which is defined in $g(\Delta)$ (projected set of mixed strategies) and which converges point-wise to fixed points, we can define $\textrm{acp}_{\textrm{sc}},\textrm{acp}_{\textrm{sw}}$ to be the average case performance as in Section \ref{sec:measure}.
For cost/utility functions the average price of anarchy is defined as follows:
$$
\text{APoA} \!=\!
\displaystyle\frac
{\textrm{acp}_{\textrm{sc}}}
{\min\nolimits_{\vec{s}^{\ast} \in \times_i S_i} \textrm{sc}(\vec{s}^{\ast})},\;\;
\text{APoA} \!=\!
\displaystyle\frac
{\max\nolimits_{\vec{s}^{\ast} \in \times_i S_i} \textrm{sw}(\vec{s}^{\ast})}
{\textrm{acp}_{\textrm{sw}}}.
$$

\begin{remark}
The definition of APoA does not rely on the fact that the games are congestion or network coordination nor does it rely on replicator dynamics. Its only requirements is that given a game we apply a dynamic that converges point-wise for all initial mixed strategies. Essentially APoA is a scaled version of the average performance. In the next section we show that replicator dynamics converges point-wise for congestion and network coordination games and also that the fixed points (of replicator on these classes of games) with region of attraction of positive measure are Nash equilibria. In particular APoA is well-defined.
\end{remark} 

%\bigskip

\section{Analysis of Replicator Dynamics in Potential Games}
\label{sec:tools}

In this section we develop the mathematical machinery necessary for computing the average case performance of replicator dynamics in
different classes of potential games. Specifically, we establish point-wise convergence of replicator dynamics for linear congestion games and arbitrary networks of coordination games  (Theorem \ref{thm:congestionpointwise}). This allows us to define properly the average case performance which is essentially equal to the weighted sum of the social cost/welfare of all equilibria weighted by the cumulative measure/volume of all initial conditions that converge to each (point-wise). Next, we show that the union of regions of attraction of (locally) unstable equilibria is of measure zero (Theorem \ref{zero}). Combining this result with a game theoretic characterization of (un)stable equilibria in  \cite{Kleinberg09multiplicativeupdates}, known as weakly stable equilibria, establishes that only weakly stable equilibria affect the average case system performance. The analysis here is a strengthening of the techniques of \cite{Kleinberg09multiplicativeupdates}
to carefully account for the possibility of continuums of unstable equilibria. Finally, we still need to compute for each weakly stable equilibrium the size of its region of attraction. The tool that is necessary for this is to establish invariants for replicator dynamics in different classes of games. We present an information theoretic invariant function (Theorem \ref{lemma:KL}) for replicator dynamics for bipartite network coordination games.  Such invariant functions have been identified \cite{PiliourasAAMAS2014, Soda14} for network extensions of zero sum games \cite{dask09,Cai}.

\subsection{Poinwise Convergence}\label{sec:pointwise}
We show that replicator dynamics converges point-wise for the class of linear congestion and network coordination games. The proof of the theorem has two steps. The first step is standard, utilizes the potential function of the game and establishes convergence to equilibria sets. The critical, second step is to construct a \emph{local} Lyapunov function in some small neighborhood of a limit point.
%Remark \ref{rem:isolated} argues that we can prove pointwise convergence for any congestion game that satisfies the property that the fixed points are \emph{isolated}.

\begin{theorem}\label{thm:congestionpointwise}
%For any (linear congestion)/(network coordination) game
Given any initial condition replicator dynamics converges to a fixed point (point-wise convergence) in all linear congestion and network coordination games.
\end{theorem}

\begin{proof}
We will prove here the result in the case of linear congestion games. The argument for network coordination games follows similar lines and is in the appendix \ref{appendix:congestionpointwise}.

%\begin{comment}
We denote by  $\hat{c}_{i}$ the expected cost of agent $i$ under mixed strategy profile $\vec{p}$.
Moreover, $c_{i\gamma}$ is his expected cost when he deviates to strategy $\gamma$ and all other agents still play according to $\vec{p}$.
We observe that
$%\begin{align*}
\Psi (\vec{p}) = \sum_{i} \hat{c}_{i}+ \sum_{i,\gamma}\sum_{e\in \gamma}(b_{e}+a_{e})p_{i\gamma}
$ %\end{align*}
  is a Lyapunov function since
\begin{align*}
\frac{\partial \Psi}{\partial p_{i\gamma}} &=  c_{i\gamma} + {\sum_{j \neq i}p_{j\gamma'}\frac{\partial c_{j\gamma'}}{\partial  p_{i\gamma}}+\sum_{e\in \gamma}(b_{e}+a_{e})} = %\\&=
c_{i\gamma} + \underbrace{\sum_{j \neq i,\gamma'}\sum_{e \in \gamma \cap \gamma '}a_{e}p_{j\gamma '}+\sum_{e\in \gamma}(b_{e}+a_{e})}_{c_{i\gamma}} = 2c_{i\gamma}
\end{align*}
and hence
$
\frac{d \Psi}{dt} = \sum_{i,\gamma}\frac{\partial \Psi}{\partial p_{i\gamma}}\frac{d p_{i\gamma}}{dt} = -\sum_{i,\gamma,\gamma'}p_{i\gamma}p_{i\gamma'}(c_{i\gamma}-c_{i\gamma'})^2 \leq 0
$,
 with equality at fixed points. Hence (as in \cite{Kleinberg09multiplicativeupdates}) we have convergence to equilibria sets (compact connected sets consisting of fixed points).
 Next, we will  argue that each trajectory has a unique (equilibrium) limit point.

 %This doesn't suffice for point-wise convergence.
 % To be exact it suffices only in the case the equilibria are discrete (which is not the case for linear congestion games - see lemma ~\ref{infinite1} ).\\
 %\end{comment}

Let $\vec{q}$ be a limit point of the trajectory $\vec{p} (t)$. Wlog we can assume that $\vec{p} (0)$  is in the interior of $\Delta$  and hence $\vec{p} (t)$ is in the interior of $\Delta$ for all $t \in \mathbb{R}$  (we can assume that we start in the interior of $\Delta$ otherwise we can just  consider the subgame defined by the strategies that agents play with positive probability.).
We have that $\Psi(\vec{q})\leq\Psi(\vec{p} (t))$ where the equality holds only if we start at equilibrium. We define the relative entropy $I(\vec{p}) = -\sum_{i}\sum_{\gamma : q_{i\gamma}>0}q_{i\gamma} \ln (p_{i\gamma}/q_{i\gamma}) \geq 0 \textrm{ (Jensen's inequality)}$
and $I(\vec{p})=0$ iff $\vec{p} = \vec{q}$.
We denote by  $\hat{d}_{i},d_{i\gamma}$ the expected costs of agent $i$ under the mixed strategy profile $\vec{q}$.

 %We get that
\begin{align*}
\frac{d I}{dt} &= -\sum_{i}\sum_{\gamma : q_{i\gamma}>0}q_{i\gamma}(\hat{c}_{i}-c_{i\gamma}) %\\
%&= -\sum_{i} \hat{c}_{i} + \sum_{i}\sum_{\gamma: q_{i\gamma}>0}q_{i\gamma}c_{i\gamma}\\
%&
 = -\sum_{i} \hat{c}_{i} + \sum_{i,\gamma}q_{i\gamma}c_{i\gamma}
%\\&=
%-\sum_{i} \hat{c}_{i} + \sum_{i}\sum_{\gamma}q_{i\gamma}[\sum_{j \neq i}\sum_{\gamma'}\sum_{e \in \gamma \cap \gamma %'}a_{e}p_{j\gamma '}+\sum_{e\in \gamma}(b_{e}+a_{e})]\\
\\&=
-\sum_{i} \hat{c}_{i} + \sum_{i,\gamma}\sum_{e\in \gamma}(b_{e}+a_{e})q_{i\gamma} + \sum_{i,\gamma}\sum_{j \neq i}\sum_{\gamma'}\sum_{e \in \gamma \cap \gamma '}a_{e}q_{i\gamma}p_{j\gamma '}
\\&=
-\sum_{i} \hat{c}_{i} + \sum_{i,\gamma}\sum_{e\in \gamma}(b_{e}+a_{e})q_{i\gamma} + \sum_{i,\gamma}\sum_{j \neq i}\sum_{\gamma'}\sum_{e \in \gamma \cap \gamma '}a_{e}q_{j\gamma'}p_{i\gamma }
%\\&=
%-\sum_{i} \hat{c}_{i} + \sum_{i}\sum_{\gamma}\sum_{e\in \gamma}(b_{e}+a_{e})q_{i\gamma} + %\sum_{i}\sum_{\gamma}p_{i\gamma}(d_{i\gamma} - \sum_{e\in \gamma}b_{e}+a_{e})
\\&=
-\sum_{i} \hat{c}_{i} + \sum_{i,\gamma}\sum_{e\in \gamma}(b_{e}+a_{e})q_{i\gamma} - \sum_{i,\gamma}\sum_{e\in \gamma}(b_{e}+a_{e})p_{i\gamma}+ \sum_{i,\gamma}p_{i\gamma}(d_{i\gamma})\\&=
\sum_{i}\hat{d}_{i}-\sum_{i} \hat{c}_{i} + \sum_{i,\gamma}\sum_{e\in \gamma}(b_{e}+a_{e})q_{i\gamma} - \sum_{i,\gamma}\sum_{e\in \gamma}(b_{e}+a_{e})p_{i\gamma}- \sum_{i,\gamma}p_{i\gamma}(\hat{d}_{i}-d_{i\gamma})
%\\&=\sum_{i}\hat{d}_{i} + \sum_{i}\sum_{\gamma}\sum_{e\in \gamma}(b_{e}+a_{e})q_{i\gamma} - \sum_{i} \hat{c}_{i} - %\sum_{i}\sum_{\gamma}\sum_{e\in \gamma}(b_{e}+a_{e})p_{i\gamma}- \sum_{i}\sum_{\gamma}p_{i\gamma}(\hat{d}_{i}-d_{i\gamma})
\\&= \Psi(\vec{q}) - \Psi(\vec{p})- \sum_{i,\gamma}p_{i\gamma}(\hat{d}_{i}-d_{i\gamma})
\end{align*}
The rest of the proof follows in a similar way to \cite{akin}.\\

We break the term $\sum_{i,\gamma}p_{i\gamma}(\hat{d}_{i}-d_{i\gamma})$ to positive and negative terms (the zero terms can be ignored), i.e., $\sum_{i,\gamma}p_{i\gamma}(\hat{d}_{i}-d_{i\gamma}) = \sum_{i,\gamma: \hat{d}_{i} >d_{i\gamma}}p_{i\gamma}(\hat{d}_{i}-d_{i\gamma}) + \sum_{i,\gamma: \hat{d}_{i} <d_{i\gamma}}p_{i\gamma}(\hat{d}_{i}-d_{i\gamma})$
\begin{claim}\label{claim:exist} There exists an $\epsilon >0$ so that the function $Z(\vec{p}) = I(\vec{p})+ 2\sum_{i,\gamma: \hat{d}_{i}<d_{i\gamma}}p_{i,\gamma}$ has $\frac{dZ}{dt}<0$ for $\norm[1]{\vec{p}-\vec{q}}<\epsilon$ and $\Psi(\vec{q})<\Psi(\vec{p})$.
 \end{claim}

 To prove this claim, first assume that $\vec{p} \to \vec{q}$. We get $\hat{c}_{i} - c_{i\gamma} \to \hat{d}_{i}-d_{i\gamma}$ for all $i,\gamma$. Hence for small enough $\epsilon>0$ with $\norm[1]{\vec{p}-\vec{q}}<\epsilon$, we have that $\hat{c}_{i}-c_{i\gamma} \leq \frac{3}{4} (\hat{d}_{i}-d_{i\gamma})$ for the terms which $\hat{d}_{i}-d_{i\gamma}<0$. Therefore
\begin{align*}
\frac{dZ}{dt} &=  \Psi(\vec{q}) - \Psi(\vec{p}) - \!\!\!\!\!\sum_{i,\gamma: \hat{d}_{i} >d_{i\gamma}}p_{i\gamma}(\hat{d}_{i}-d_{i\gamma}) - \!\!\!\!\!\sum_{i,\gamma: \hat{d}_{i} <d_{i\gamma}}p_{i\gamma}(\hat{d}_{i}-d_{i\gamma}) + 2\!\!\!\!\!\sum_{i,\gamma: \hat{d}_{i} <d_{i\gamma}}p_{i\gamma}(\hat{c}_{i}-c_{i\gamma})
\\&\leq \Psi(\vec{q}) - \Psi(\vec{p}) - \!\!\!\!\!\sum_{i,\gamma: \hat{d}_{i} >d_{i\gamma}}p_{i\gamma}(\hat{d}_{i}-d_{i\gamma}) - \!\!\!\!\!\sum_{i,\gamma: \hat{d}_{i} <d_{i\gamma}}p_{i\gamma}(\hat{d}_{i}-d_{i\gamma}) + 3/2 \!\!\!\!\! \sum_{i,\gamma: \hat{d}_{i} <d_{i\gamma}}p_{i\gamma}(\hat{d}_{i}-d_{i\gamma})
\\&=
\underbrace{\Psi(\vec{q}) - \Psi(\vec{p})}_{< 0} + \underbrace{\sum_{i,\gamma: \hat{d}_{i} >d_{i\gamma}}-p_{i\gamma}(\hat{d}_{i}-d_{i\gamma})}_{\leq 0} + 1/2\underbrace{\sum_{i,\gamma: \hat{d}_{i} <d_{i\gamma}}p_{i\gamma}(\hat{d}_{i}-d_{i\gamma})}_{\leq 0}<0
\end{align*}
where we substitute $\frac{p_{i\gamma}}{dt} = p_{i\gamma}(\hat{c}_{i}-c_{i\gamma})$ (replicator equations), and the claim is proved.
Note that $Z(\vec{p}) \geq 0$ (sum of non-negative terms and $I(\vec{p}) \geq 0$) and is zero iff $\vec{p} = \vec{q}. \;\; \textrm{(i)}$

%\skip

To finish the proof of the theorem, let $\vec{q}$ be a non-trivial limit point of $\vec{p} (t)$ (i.e., $\vec{p} (0)$ is not a fixed point). There exists an increasing sequence of times $t_{n}$, with $t_{n} \to \infty$ and $\vec{p} (t_{n}) \to \vec{q}$. We consider $\epsilon '$ such that the set $C = \{\vec{p} : Z(\vec{p})< \epsilon '\}$ is inside $B = \norm[1]{\vec{p}-\vec{q}}<\epsilon$ where $\epsilon$ is from claim above. Since $\vec{p}( t_{n}) \to \vec{q}$, consider a time $t_{N}$ where $\vec{p} (t_{N})$ is inside $C$. From the claim above we get that $Z(\vec{p})$ is decreasing inside $B$ (and hence inside $C$), thus $Z(\vec{p} (t)) \leq Z(\vec{p} (t_{N})) < \epsilon '$ for all $t \geq t_{N}$, hence the orbit will remain in $C$. By the fact that $Z(\vec{p} (t))$ is decreasing in $C$ (claim above) and also $Z(\vec{p} (t_{n})) \to Z(\vec{q}) =0$ it follows that $Z(\vec{p} (t)) \to 0$ as $t \to \infty$. Hence $\vec{p} (t) \to \vec{q}$ as $t \to \infty$ using (i).
\end{proof}

\begin{remark}\label{rem:isolated} If the fixed points of the dynamics are \emph{isolated} then a (global) Lyapunov function suffices to show that the system converges point-wise (first step of the proof above). A fixed point $\vec{x}_0$ is called isolated, if there exists an neighborhood of $\vec{x}_0$ so that $\vec{x}_0$ is the unique fixed point in that neighborhood. However, this is not the case even in linear congestion games (see Lemma \ref{lem:infinite1} for examples of linear congestion games with continuums of (Nash) equilibria).
\end{remark}
%The same result can be shown for network coordination games. The proof is almost identical and can be found in the appendix.
%\begin{theorem}\label{thm:networkpointwise}
%For any network coordination game, replicator dynamics converges to a fixed point (point-wise convergence).
%\end{theorem}

%The proof is almost the same as the proof of theorem \ref{congestionpointwise} and can be found in the appendix.

\subsection{Global Stability Analysis}\label{sec:stability}

Replicator dynamics  in linear congestion games and network coordination games (as well as  any dynamic that converges point-wise) induces a probability distribution over the fixed points. The probability assigned to each fixed point is proportional to the volume of its region of attraction. The fixed points can be exponentially many or even accountable many, but as it is stated below (corollary \ref{col:weakly}), only the weakly stable Nash equilibria have non-zero volumes of attraction.

\begin{definition}\cite{Kleinberg09multiplicativeupdates} A Nash equilibrium is called weakly stable if given any two randomizing agents, fixing one of the agents to choosing one of his strategies with probability one, leaves the other agent indifferent between the strategies in his support. That is a Nash equilibrium \textbf{p} is weakly stable if for any agents $i,j$ and strategies $\gamma,\gamma' \in S_{i},\delta \in S_{j}$ with $p_{i\gamma},p_{i\gamma'},p_{j\delta}>0$: $c_i(\gamma,\delta,\textbf{p}_{-ij})=c_i(\gamma',\delta,\textbf{p}_{-ij})$.
\end{definition}

 {\cite{Kleinberg09multiplicativeupdates}}  showed that in congestion games every stable fixed point is a weakly stable Nash equilibrium. The following theorem (that assumes point-wise convergence) has as a corollary that for all but a measure zero set of initial conditions replicator dynamics converges to a weakly stable Nash equilibrium.

\begin{theorem}\label{zero} The set of initial conditions for which the replicator converges to unstable fixed points has measure zero in $\Delta$ for linear congestion games and network coordination games.
\end{theorem}

\textit{Sketch.} The proof of this theorem relies on dedicated machinery from topology and dynamical systems theory. These tools and the complete proof are presented in detail in the appendix \ref{appendix:zero}.
The main conceptual steps are as follows: First, since the space of mixed strategy profiles (i.e., products of simplices) are of zero measure in their native space we work with projections on subspaces where the set of initial conditions has full measure.  Due to  a classic theorem in dynamical systems (Center-Stable Manifold theorem) we have that the set of initial conditions that stay trapped in a small enough neighborhood of an unstable equilibrium is a zero measure set. Any initial condition that converges (pointwise) to this unstable fixed point must (at some time $t$) reach points in this set. All of these initial conditions can thus be covered by a countable union of preimages of the zero measure neighborhood implied by the Center-Stable Manifold theorem. Due to the smoothness of the flow (a technical condition known as diffeomorphism) these preimages must also be of zero measure and the countable union of zero measure sets imply a zero measure region of attraction for each unstable equilibrium. The only remaining hurdle is the case where the game has continuum of equilibria. In this case, although the region of attraction of each equilibrium is of zero measure, their union could have positive measure. Due to compactness of state space, we argue that it suffices to cover each (unstable) equilibrium set with a finite cover of (zero-measure) neighborhoods. At this point standard union bound arguments suffice to complete the argument. \qed

%\medskip
This theorem extends to all congestion games for which the replicator dynamics converges point-wise (e.g., games with finite equilibria).
Combining theorem \ref{zero} with the weakly stable characterization of \cite{Kleinberg09multiplicativeupdates} which holds for all congestion/potential games, we get the following:

\begin{comment}
\begin{theorem} \label{stableweakly} \cite{Kleinberg09multiplicativeupdates} In replicator dynamics on congestion games and network coordination games, every stable fixed point is a weakly stable Nash equilibrium.
\end{theorem}
\begin{proof} In \cite{Kleinberg09multiplicativeupdates}, the statement is proved for congestion games. Since a network coordination game is a potential game and every potential game is isomorphic to a congestion game, the rest follows.
\end{proof}
Combining theorem \ref{zero} with theorem \ref{stableweakly} we have the following corollary:
\end{comment}

\begin{corollary}
\label{col:weakly}
In linear congestion games and network coordination games,
for all but a measure zero set of initial conditions, replicator dynamics converges (point-wise) to weakly stable Nash equilibria.\end{corollary}

\subsection{Invariant Functions from Information Theory}
\label{sec:invariants}

We have established that all attracting (i.e., asymptotically stable) fixed points are weakly stable Nash equilibria.
We still need to characterize and compute the regions of attraction of these equilibria.
The key idea here is to characterize the boundaries of the regions of attraction.
This is due to the following theorem:

%Finding bounds or computing average PoA is not an easy task in most of the cases. Computing average PoA essentially means to learn the probability distribution over the fixed points (as it occurs by the dynamics, in point-wise convergent systems), i.e compute the regions of attraction of the fixed points.
\begin{theorem}\label{boundary}\cite{khalil} If $\vec{q}$ is an asymptotically stable equilibrium point for a system $\dot{x} = f(x)$ where $f \in C^1$, then its region of attraction $R_{\vec{q}}$ is an invariant set whose boundaries are formed by trajectories.
\end{theorem}

If we identify a (continuous) invariant function $f$, \textit{i.e.}, a function that remains constant on any trajectory, and $\vec{q}$ is a (limit) point of the trajectory then the whole trajectory lies on the set $\{\vec{x}:f(\vec{x})=f(\vec{q})\}$. If we identify more invariant functions $f_1,f_2,\dots, f_k$
then the whole trajectory lies on the set $\{\vec{x}:f_1(\vec{x})=f_1(\vec{q}) \wedge f_2(\vec{x})=f_2(\vec{q}) \wedge   \dots \wedge f_k(\vec{x})=f_k(\vec{q}) \}$. By identifying enough invariant functions, we can derive an exact algebraic description of the trajectory.

By our point-wise convergence result each trajectory converges to an equilibrium.
So each point of the state space that does not belong in the region of attraction of a weakly stable equilibrium, must converge to an unstable equilibrium. By computing the (union of) regions of attraction of all unstable equilibria we can understand how they partition the state space into regions of attractions for the asymptotically stable equilibria\footnote{The region of attraction of an unstable equilibrium is referred to as the stable manifold of the (unstable) fixed point.}. All points on the stable manifold of unstable fixed point $\vec{q}$  lie on the set $\{\vec{x}:f_1(\vec{x})=f_1(\vec{q}) \wedge f_2(\vec{x})=f_2(\vec{q}) \wedge   \dots \wedge f_k(\vec{x})=f_k(\vec{q}) \}$ where $f_1, \dots, f_k$ the invariant functions of the dynamic. Such descriptions can allow for exact computation of volumes
 of regions of attraction (Section \ref{sec:staghunt}), approximate volume computation (Section \ref{sec:APoA2x2}), designing efficient oracles for testing if an initial condition belongs to the region of attraction of an equilibrium (Section \ref{sec:star}), and computing average system performance, amongst other applications.

The following lemma that identifies invariants functions in bipartite coordination games follows straightforwardly from prior work on identifying invariant functions for network generalizations of (linear transformations of) zero-sum games \cite{PiliourasAAMAS2014, Soda14}).
To prove any such statement it suffices to compute the time derivatives of these functions along any trajectory and show them to be equal to zero. %For completeness, we provide the proof in appendix \ref{appendix:KL}.

\begin{comment}
Assume that we have an oracle function $f$ that can be computed-approximated in polynomial time such that for every (almost every) initial condition $\vec{p}_{0}$, $f(\vec{p}_{0})$ gives us the fixed point the dynamic will eventually converge to. Then by simply taking "enough" (polynomial in size) samples, we can have a good estimation for the volume of regions of attraction of the fixed points. So finding an efficient oracle, gives us an efficient (through sampling) algorithm for computing average PoA. The hardest and less trivial part is to find such an efficient oracle $f$.
\end{comment}

\begin{lemma}
\label{lemma:KL}
Let $\vec{p}(t) = (\vec{p}_{1}(t),...,\vec{p}_{N}(t))$  be a trajectory of replicator dynamics when applied to a bipartite network of coordination games that has a fully mixed Nash equilibrium $\vec{q} = (\vec{q}_{1},...,\vec{q}_{N})$ then the function $\sum_{i \in V_{left}}  H(\vec{q}_{i}, \vec{p}_{i}(t))-\sum_{i \in V_{right}}  H(\vec{q}_i, \vec{p}_{i}(t))$ is (time-)invariant, where $H(\vec{x},\vec{y})= -\sum_i x_i\ln y_i$.
\end{lemma}

The cross entropy between the Nash $\vec{q}$ and the state of the system, however is equal to the summation of the K-L divergence  between these two distributions and the entropy of $\vec{q}$. Since the entropy of $\vec{q}$ is constant, we derive the following corollary (rephrasing the previous lemma):

 \begin{corollary}
\label{coro1}
Let $\vec{p}(t)$  be a trajectory of the replicator dynamic when applied to a bipartite network of coordination games that has a fully mixed Nash equilibrium $\vec{q}$ then the K-L divergence between $\vec{q}$ and the $\vec{p}(t)$ is constant, \textit{i.e.}, does not depend on $t$.
\end{corollary}

%\bigskip

%\section{Average Case Analysis and Bounds}

%This is the average case analysis.

%\subsection{Coordination Games}\label{sec:averagecoordination}

%\subsection{Load Balancing Games}\label{sec:averageload}

%\newpage

%In this section, we exploit the tools and tech

\section{Applications of Average Case Analysis}
\label{sec:average}

We will use the tools we have developed in the previous section to compute the regions of attractions  and find the average case performance of replicator dynamics in classic game theoretic settings. The games we examine are: the Stag Hunt game, (parametric) coordination games, polymatrix coordination games played over a star as well as symmetric linear load balancing games.

\subsection{Exact Quantitative Analysis of Risk Dominance in the Stag Hunt Game}
\label{sec:staghunt}

\begin{figure}
\hfill\hfill
\subfigure[Stag Hunt game]
{
\centering
\begin{footnotesize}
\begin{game}{2}{2}[][][]
        \> Stag\> Hare   \\
Stag      \> 5, 5  \> 0, 4  \\
Hare    \> 4, 0  \> 2, 2    \\
\end{game}
\end{footnotesize}
\label{fig:game1}
}
\hfill\hfill
\subfigure[$w$-coordination game]
{
\centering
\begin{footnotesize}
\begin{game}{2}{2}[][][]
        \> Stag \> Hare   \\
Stag      \> 1, 1  \> 0, 0  \\
Hare    \> 0, 0  \> w, w   \\
\end{game}
\end{footnotesize}
\label{fig:game2}
}
\hfill
\hfill
\hfill
\caption{}
\end{figure}

\begin{comment}
\begin{figure}[h!]
\begin{center}
\begin{footnotesize}
\begin{game}{2}{2}[][][]
        \> Heads \> Tails   \\
Heads      \> 1, 1  \> 0, 0  \\
Tails    \> 0, 0  \> 2, 2    \\
\end{game}
\end{footnotesize}
\end{center}
\begin{caption}{A coordination game which is a best response equivalent to Stag Hunt.}\end{caption}
\end{figure}

\begin{figure}
  \centering
  \includegraphics[width=.75\linewidth]{stag.png}
\caption{a) Stag Hunt game - b) A coordination game which is a best response equivalent to Stag Hunt.}
\label{fig:game1}
\end{figure}
\end{comment}

\begin{comment}

\subfigure[Stag Hunt game.]
{
\centering
\begin{footnotesize}
\begin{game}{2}{2}[][][]
        \> Stag\> Hare   \\
Stag      \> 5, 5  \> 0, 4  \\
Hare    \> 4, 0  \> 2, 2    \\
\end{game}
\end{footnotesize}
\label{fig:game1}
}
\hfill\hfill
\subfigure[$w$-coordination game]
{
\centering
\begin{footnotesize}
\begin{game}{2}{2}[][][]
        \> Stag \> Hare   \\
Stag      \> 1, 1  \> 0, 0  \\
Hare    \> 0, 0  \> w, w   \\
\end{game}
\end{footnotesize}
\label{fig:game2}
}
\hfill
\hfill
\hfill
\caption{}
\end{comment}

\begin{figure}
  \centering
  \includegraphics[width=.35\linewidth]{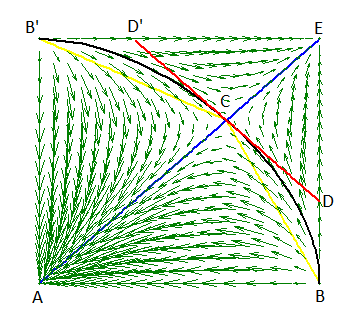}
  \caption{Vector field of replicator dynamics in Stag Hunt.}
  \label{fig:game13}
\end{figure}

\begin{comment}
\begin{figure}[h!]
\begin{center}
\begin{footnotesize}
\begin{game}{2}{2}[][][Left]
        \> Left \>  Right \\
Left \> 1, 1, 1     \> 2/3, 2/3, 2/3  \\
Right    \> 2/3, 2/3, 2/3     \> 2/3, 2/3, 2/3  \\
\end{game}\hspace{.5in}
\begin{game}{2}{2}[][][Right]
             \> Left   \> Right   \\
Left  \> 2/3, 2/3, 2/3 \> 2/3, 2/3, 2/3  \\
Right     \> 2/3, 2/3, 2/3  \> 1, 1, 1 \\
\end{game}
\end{footnotesize}
\end{center}
\end{figure}
\end{comment}

The Stag Hunt game (figures \ref{fig:game1}) has two pure Nash, $(Stag, Stag)$ and $(Hare, Hare)$ and a symmetric mixed Nash equilibrium where each agent chooses strategy $Hare$
 with probability $2/3$.
Stag Hunt replicator trajectories are equivalent those of a coordination game with $w=2$.\footnote{If both agents reduce their payoff of their first strategy by $4$, the replicator trajectories remain invariant. This results to a $w$-coordination game with $w=2$.} Coordination  games are potential games where the potential function in each state is equal to the utility of each agent.
Since the mixed Nash is not weakly stable
replicator dynamics  converges to pure Nash
equilibria for all but a zero measure of initial conditions (Theorem \ref{zero}). When we study the replicator dynamic here, it suffices to examine its projection in the subspace $p_{1s}\times p_{2s}\subset (0,1)^2$
which captures the evolution of the probability that each agent assigns to strategy $Stag$ (see figure \ref{fig:game13}).
Using the invariant property of lemma \ref{lemma:KL}, we compute
the size of each region of attraction in this space and thus provide a quantitative analysis of risk dominance in the classic Stag Hunt game.

\begin{theorem}
\label{thm:SH}
%The replicator flow $\phi$, when applied to the Stag Hunt game has two equilibria with regions of attraction of positive Lebesgue measure.
%These correspond to the two pure Nash equilibria of the game $(Stag, Stag)$ and $(Hare, Hare)$.
The region of attraction of $(Hare, Hare)$ is the subset
of $(0,1)^2$ % the interior of the state space
that satisfies   $p_{2s}<\frac12 (1 - p_{1s} + \sqrt{1 + 2 p_{1s} - 3 p_{1s}^2})$ and has Lebesgue measure
$\frac{1}{27} (9 + 2 \sqrt{3} \pi) \approx 0.7364$. The region of attraction of $(Stag, Stag)$ is the subset
of  $(0,1)^2$ %the interior of the simplex
that satisfies   $p_{2s}>\frac12 (1 - p_{1s} + \sqrt{1 + 2 p_{1s} - 3 p_{1s}^2})$ and has Lebesgue measure
$\frac{1}{27} (18 - 2 \sqrt{3} \pi) \approx 0.2636$. The stable manifold of the mixed Nash equilibrium satisfies the equation
$p_{2s}=\frac12 (1 - p_{1s} + \sqrt{1 + 2 p_{1s} - 3 p_{1s}^2})$ and has zero Lebesgue measure.
\end{theorem}

%\begin{proof} See appendix.
%\end{proof}

\begin{proof}
%Since Stag Hunt is payoff equivalent to a  coordination game and has a fully mixed Nash equilibrium, %we have that %for each
  In the case of Stag Hunt games, one can verify in a straightforward manner (via substitution)  that
  $\frac{d\big(\frac{2}{3}\ln(\phi_{1s}(t,\vec{p}))+  \frac{1}{3}\ln(\phi_{1h}(t,\vec{p})) -   \frac{2}{3}\ln(\phi_{2s}(t,\vec{p})) - \frac{1}{3}\ln(\phi_{2h}(t,\vec{p}))\big)}{dt}= 0$, where $\phi_{i\gamma}(t,\vec{p})$, corresponds to the probability
  that each agent $i$ assigns to strategy $\gamma$ at time $t$ given initial condition $\vec{p}$.
  %(from corollary~\ref{coro1}) .
  This is a special case of corollary~\ref{coro1}.
  We use this invariant function to identify the stable and unstable manifold of the interior Nash $\vec{q}$.

 Given any point $\vec{p}$ of the stable manifold of $\vec{q}$, we have that by definition $\lim_{t\rightarrow \infty} \phi(t,\vec{p}) = \vec{q}$.
 Similarly for the unstable manifold, we have that $\lim_{t\rightarrow -\infty} \phi(t,\vec{p}) = \vec{q}$. The time-invariant property implies that
 for all such points (belonging to the stable or unstable manifold), $\frac{2}{3}\ln(p_{1s})+\frac{1}{3}\ln(1-p_{1s})$$-\frac{2}{3}\ln(p_{2s})-\frac{1}{3}\ln(1-p_{2s})=
 \frac{2}{3}\ln(q_{1h})+\frac{1}{3}\ln(1-q_{1h})$$-\frac{2}{3}\ln(q_{2h})-\frac{1}{3}\ln(1-q_{2h})=0$, since the fully mixed Nash equilibrium is symmetric.
This condition is equivalent to  $p^2_{1s}(1-p_{1s})=p^2_{2s}(1-p_{2s})$, where $0<p_{1s}, p_{2s}<1$.
 It is straightforward to verify that
 this algebraic equation is satisfied by the following two distinct solutions, the diagonal line $(p_{2s}=p_{1s})$ and $p_{2s}=\frac12 (1 - p_{1s} + \sqrt{1 + 2 p_{1s} - 3 p_{1s}^2})$.
  Below, we show that these manifolds correspond indeed to the state and unstable manifold of the mixed Nash, by showing that this Nash
  equilibrium satisfies these equations and by establishing that the vector field is tangent everywhere along them.

  The case of the diagonal is trivial and follows from the symmetric nature of the game. We verify the claims about $p_{2s}=\frac12 (1 - p_{1s} + \sqrt{1 + 2 p_{1s} - 3 p_{1s}^2})$.
  Indeed, the mixed equilibrium point in which $p_{1s}=p_{2s}=2/3$ satisfies the above equation.
 %\dots
 We establish that the vector filed is tangent to this manifold
 by showing in Lemma \ref{lemma:tangent} that $\frac{\partial p_{2s}}{\partial p_{1s}}= \frac{\frac{dp_{2s}}{dt}}{\frac{dp_{1s}}{dt}}\defeq\frac{p_{2s}\big(u_2(s)-(p_{2s}u_2(s)+(1-p_{2s})u_2(h)) \big)}{p_{1s}\big(u_1(s)-(p_{1s}u_1(s)+(1-p_{1s})u_1(h))\big)}$, where the last equality is derived by the definition of replicator dynamics. Finally, this manifold is indeed attracting to the equilibrium. Since the function $p_{2s}=y(p_{1s})=\frac12 (1 - p_{1s} + \sqrt{1 + 2 p_{1s} - 3 p_{1s}^2})$ is a strictly decreasing function of $p_{1s}$ in [0,1] and satisfies $y(2/3)=2/3$, this implies that its graph is contained in the subspace $\big(0<p_{1s}<2/3 \cap 2/3<p_{2s}<1\big) \cup \big(2/3<p_{1s}<1 \cap 0<p_{2s}<2/3\big)$. In each of these subsets $\big(0<p_{1s}<2/3 \cap 2/3<p_{2s}<1\big), \big(2/3<p_{1s}<1 \cap 0<p_{2s}<2/3\big)$ the replicator vector field coordinates have fixed signs that ``push'' $p_{1s}, p_{2s}$ towards their respective mixed equilibrium values.

 The stable manifold partitions the set $0<p_{1s},p_{2s}<1$ into two subsets, each of which is flow invariant since the unstable manifold itself is flow invariant.
 Our convergence analysis for the generalized replicator flow implies that in each subset all but a measure zero of initial conditions must converge to its respective pure
 equilibrium.
 The size of the lower region of attraction\footnote{This corresponds to the risk dominant equilibrium $(Hare, Hare)$.}  is equal to the following definite integral
$\int_0^1  \frac12 (1 - p_{1s} + \sqrt{1 + 2 p_{1s} - 3 p_{1s}^2}) dx = \Big[1/2 \Big(p_{1s} - \frac{p_{1s}^2}{2} + (-\frac16 + \frac{p_{1s}}{2}) \sqrt{1 + 2 p_{1s} - 3 p_{1s}^2} -
   \frac{2 arcsin[\frac12 (1 - 3 p_{1s})]}{3 \sqrt{3}}\Big) \Big]_0^1= \frac{1}{27} (9 + 2 \sqrt{3} \pi)= 0.7364$ and the theorem follows.
   %The resulting average case price of anarchy for the case of the coordination game is $1.7364$.
\end{proof}

\subsection{Average Price of Anarchy Analysis %Risk Dominance
in Coordination/Consensus Games
via Polytope Approximations of Regions of Attraction}
\label{sec:APoA2x2}

\begin{comment}
In the previous section, we used the algebraic equations implied from lemma~\ref{lemma:KL} to find an explicit description of the stable/unstable
manifold of the mixed Nash equilibrium. These in turn were used to compute exactly the size of the regions of attraction for each of the two pure Nash
equilibria. As we move away from single instance games towards classes of games, the task of finding exact explicit descriptions of the topology of the attractor landscape
 becomes infeasible quickly.
 \end{comment}

We focus on a parametric family of coordination games, as described in figure~\ref{fig:game2}.
 We denote an instance of such a game a $w$-coordination/consensus game.  We take the $w$ parameter to be greater or equal to $1$\footnote{It is easy to see that for any $0<w<1$, $w$-coordination game is isomorphic to $1/w$-coordination game after relabeling of strategies.
Also, the replicator trajectories in the $2$-coordination game are equivalent to the standard Stag Hunt game.}.
  This game captures strategic situations where agents must learn to coordinate on a single action and where one pure equilibrium (consensus outcome) is preferable for both agents. The initial condition of the replicator dynamics captures each agent's initial bias. Both agents update their beliefs/distributions by applying the replicator and eventually the system converges to an equilibrium. Interestingly, since the mixed Nash is not weakly stable, Theorem \ref{zero} implies that the agents will reach a consensus with probability $1$
  as long as the initial conditions are chosen according to an arbitrary distribution $F$ admitting a density w.r.t. the Lebesgue measure.
  A natural such prior (distribution) is the uniform one, since it encodes a total ignorance of the agents' initial biases.
  We wish to understand what is the expected system performance given a uniformly random initial condition. Although the inefficient equilibrium will arise with positive probability hopefully its probability is small enough that no matter the  $w$ efficiency gap between the two pure equilibria the average system system performance is always within an absolute constant of the optimal, independent of $w$. We will show that this is indeed the case.

\begin{comment}
For any $w$, $G(w)$ is a coordination/potential game and therefore it is  payoff equivalent to a congestion game. The only two weakly stable equilibria
are the pure ones, hence in order to understand the average case system performance it suffices to understand the size of regions of attraction for each of them.
As in the case of Stag Hunt game, we focus on the projection of the system to the subspace $(p_{1s},p_{2s})\subset [0,1]^2$. We show the following in appendix \ref{appendix:GenSH}:
\end{comment}

\begin{theorem}
\label{thm:GenSH}
The average price of anarchy of a $w$-coordination game with $w\geq 1$ is at most $\frac{w^2+w}{w^2+1}$ and at least $\frac{w(w+1)^2}{w(w+1)^2-2w+2}$.
\end{theorem}

\textit{Sketch.} For $w$-coordination games it is straightforward to see that $p^w_{1s}(1-p_{1s})-p^w_{2s}(1-p_{2s})$ is an invariant property of the replicator system (follows from lemma \ref{lemma:KL}). The presence of the parameter $w$ on the exponent precludes
 the existence of a simple, explicit, parametric description of all the solutions.  We analyze the topology of the basins of attractions and produce simple subsets/supersets polytope approximations
 of them (see figure \ref{fig:game13}). The volume of these polytope approximations  can be computed explicitly and these measures can be used to provide upper and lower bounds on the average case system
  performance and average price of anarchy.  We present the complete proof in the appendix \ref{appendix:GenSH}.
\qed
\medskip

By combining the exact analysis of the standard Stag Hunt game (theorem~\ref{thm:SH}), theorem~\ref{thm:GenSH}  and optimizing over $w$ we derive that:

\begin{corollary}
The average price of anarchy of the class of $w$-coordination games  with $w>0$ is at least $\frac{2}{1+\frac{9+2\sqrt{3}\pi}{27}}\approx 1.15$ and  at most $\frac{4+3\sqrt{2}}{4+2\sqrt{2}}\approx 1.21$. % 1.20710     %This happens at 1+sqrt(2)
In  comparison, the price of anarchy for this class of games is unbounded.

%Furthermore, the average price of anarchy is at least $1+2\frac{\sqrt{17}-1}{(3+\sqrt{17})[(7+\sqrt{17})^2/16-2]+8}\approx 1.12$  %1.12791  % This happens at  (3+sqrt(17))/4
\end{corollary}

%Our analysis essentially shows that as $w$ grows, the size of the attraction basin of the optimal equilibrium grows at a sufficiently fast pace
%to counteract the presence of any increasingly suboptimal equilibrium. %On average the system behaves within

\subsection{Coordination/Consensus Games on a Star Graph}
\label{sec:star}

In this subsection we show how to estimate the topology of regions of attraction for star networks of $w$-coordination games.
%Specifically, we focus on a star topology where each node is an agent and each edge corresponds to our benchmark example of the Stag Hunt game.
This corresponds to strategic settings where some agents again need  to reach consensus but where there is an agent who works as a center communicating with all agents at once. The price of anarchy and stability of these games remain unchanged as we increase the size of the star. Specifically the price of stability is equal to $1$ whereas the price of anarchy can become arbitrarily large for large enough $w$.
We will argue once again that the average performance is approximately optimal.

This game has two pure Nash equilibria where all agents either play the first strategy (\textit{i.e.}, $Stag$), or the second  (\textit{i.e.}, $Hare$).
%Since in the $w$-coordination game the all $Hare$ outcome is the good equilibrium, whereas the all $Stag$ outcome is the bad one, sometimes we abbreviate strategies and use $A$ ($i.e.$, good strategy) for $Hare$ and $B$ ($i.e.$, bad strategy) for $Stag$.
For simplicity in notation sometimes we denote the first strategy, \textit{i.e.}, $Stag$, as strategy $A$ and the other strategy, \textit{i.e.,} $Hare$, as strategy $B$.
This game has a continuum of mixed Nash equilibria. Our goal is to produce an oracle which given as input an initial condition outputs the resulting equilibrium that system converges to.

\begin{comment}
\begin{figure}
\centering
\begin{subfigure} %{.5\linewidth} %\textwidth
  \centering
  \includegraphics[width=.17\linewidth]{star_2_left}   % used to 10      % used to be 5
  \caption{Examples of stable manifolds for star coordination game with $3$ agents.}
  \label{fig:sub1}
\end{subfigure}%
\begin{subfigure} %{.5\linewidth}
  \centering
  \includegraphics[width=.17\linewidth]{star_2_right}  % used to be 9    % used to be 7
  \caption{Stable manifolds lie on the intersection of level sets of invariant functions.}
  \label{fig:sub2}
\end{subfigure}
%\caption{Sample replicator trajectories in simple $2\times2$ agent games. Each point encodes the probability assigned by the agents to their first strategy.}
\label{fig:test}
\end{figure}
\end{comment}

\begin{figure}
\centering     %%% not \center
\subfigure[Examples of stable manifolds for different mixed Nash.]{\label{fig:a}\includegraphics[width=38mm]{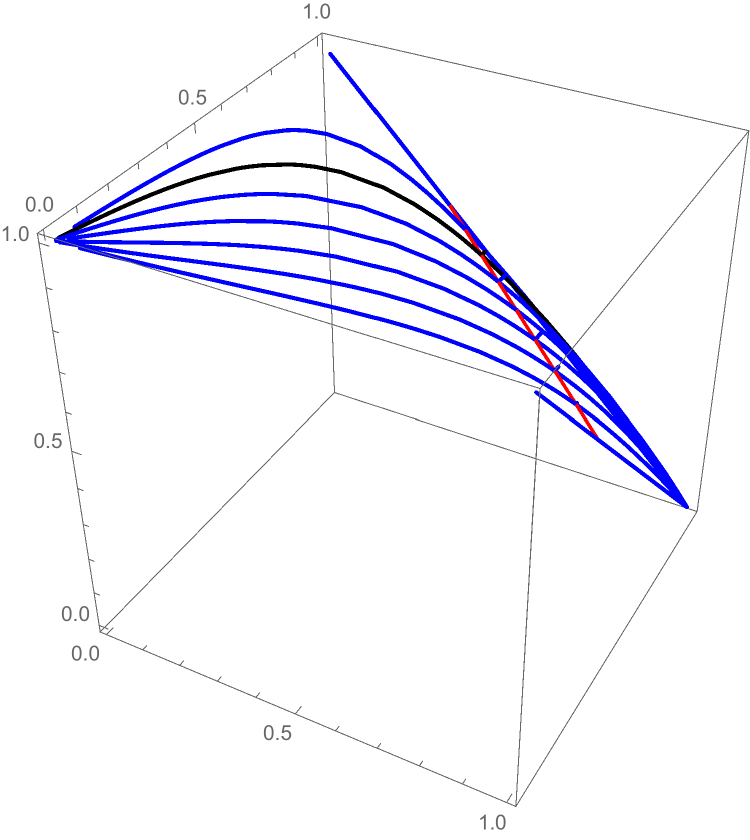}}
\subfigure[Stable manifolds lie on the intersection of level sets of invariant functions.]{\label{fig:b}\includegraphics[width=38mm]{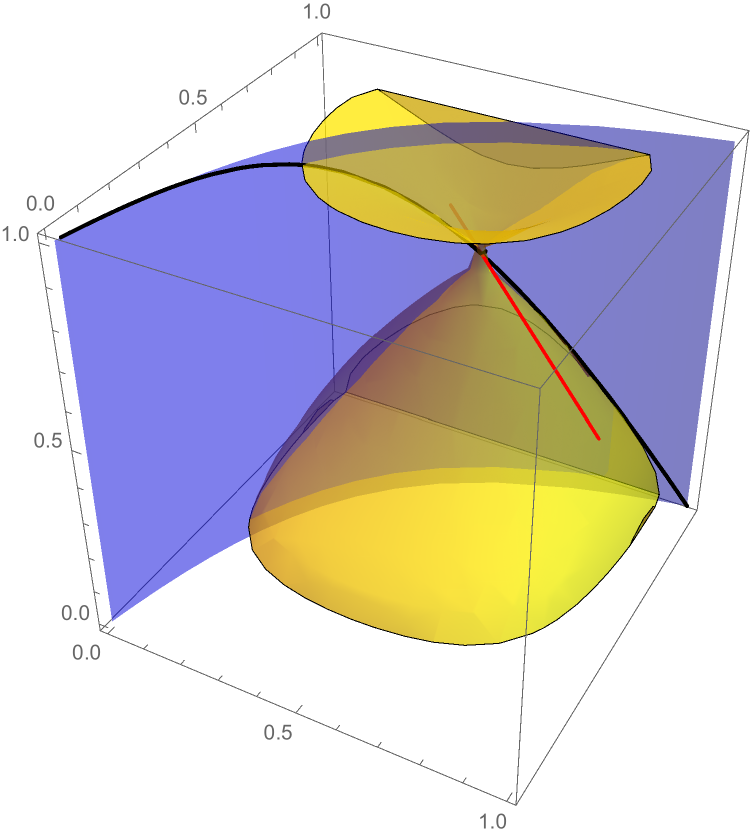}}
\caption{Star network coordination game with $3$ agents}
 \label{fig:gamestar}
\end{figure}

\begin{comment}
\begin{figure}
  \centering
  \includegraphics[width=.32\linewidth]{soda2.png} %42
  \caption{Examples of stable manifolds and invariant sets for star with $3$ agents}
  \label{fig:game4}
\end{figure}

\begin{figure}
  \centering
  \includegraphics[width=.32\linewidth]{soda3.png}  %43
  \caption{Examples of stable manifolds for star with $3$ agents}
  \label{fig:game3}
\end{figure}
\end{comment}

%Stag=A, Hare=B
\medskip

\textbf{Example.} In order to gain some intuition on the construction of these oracles let's focus on the minimal case with a continuum of equilibria ($N=3$ agents/vertices, center agent with $n=2$ neighbors).
Since each agent has two strategies it suffices to depict for each one the probability with which they choose strategy $A$ (the ``bad" $Stag$ strategy). Hence, the phase space can be depicted in $3$ dimensions. Figure \ref{fig:gamestar} depicts this phase space.
The point $(0,0,0)$ captures the good pure Nash (all $B$), whereas the point $(1,1,1)$ the bad pure Nash (all $A$).
There is also a continuum of unstable mixed Nash equilibria. Specifically, it suffices that the center player chooses $A$ with probability $w/(w+1)$ and  the summation of the probabilities that the two other agents assign to $A$ is exactly $2w/(w+1)$.
 In figure \ref{fig:gamestar}, we have chosen $w=2$. The continuum of equilibria corresponds to the red straight line.
These are unstable equilibria and by Theorem \ref{zero} almost all initial conditions are attracted to the two attracting pure Nash. For any mixed Nash equilibrium there exists a curve (co-dimension 2) of points that converge to it.
%\footnote{Even unstable equilibria have uncountably many starting points that converge to them.}.
 Figure \ref{fig:a} depicts several such stable manifolds for  sample mixed equilibria along the equilibrium line. The union of these stable manifolds partitions the state space into two regions, one attracting to equilibrium $(A, A, A)$ and the other attracting to the equilibrium $(B, B, B)$). Hence, in order to construct our oracle it suffices to have a description of these attracting curves for the mixed equilibria. However, as shown in figure \ref{fig:b}, we have identified two distinct invariant functions for the replicator dynamic in this system. Given any mixed Nash equilibrium, the set of points of the state space which agree with the value of each of these invariant functions define a set of co-dimension one (the double hollow cone and the curved plane). Any points that converge to this equilibrium must lie on the intersection of these sets (black curve). In fact, due to our point-wise convergence theorem,  it immediately follows that this intersection is exactly the stable manifold of the unstable equilibrium. The case for general $N(=n+1)$ works analogously, but now we need to identify $N-1$ ($=n$, equal to the number of neighbors) invariant functions in an algorithmic, efficient manner.

\begin{center}
\fbox{\parbox{5.5in}{
\begin{minipage}{5.4in}
\begin{tt}
{\bf Oracle}
\begin{enumerate}
\item Input: Initial condition $(x_{1}(0),\dots,x_{n}(0),y(0))$
\item Output: A or B or mixed
\item If $\sum x_{i}(0)>(\geq)\frac{w}{w+1}n$ and $y(0)\geq (>)\frac{w}{w+1}$ return A.
\item If $\sum x_{i}(0)<(\leq)\frac{w}{w+1}n$ and $y(0)\leq (<)\frac{w}{w+1}$ return B.
\item Compute by solving system \ref{system}-\ref{system2} $x_{1},\dots,x_{n}$ (binary search)
\item Let $f(\rho) = \left(\frac{\rho(w+1)}{w}\right)^{\frac{w}{w+1}}\left[(1-\rho)(w+1)\right]^{\frac{1}{w+1}}- \prod_{i} \left(\frac{x_{i}(0)}{x_{i}}\right)^{x_{i}}\left(\frac{1-x_{i}(0)}{1-x_{i}}\right)^{1-x_{i}}$
\item if ($\sum_{i}x_{i}(0)>\frac{w}{w+1}n$ and $f(y(0))<0$) or \\($\sum_{i}x_{i}(0)<\frac{w}{w+1}n$ and $f(y(0))>0$) return B.
\item if ($\sum_{i}x_{i}(0)>\frac{w}{w+1}n$ and $f(y(0))>0$) or \\($\sum_{i}x_{i}(0)<\frac{w}{w+1}n$ and $f(y(0))<0$) return A.
\item return mixed fixed point $(x_1,\dots,x_n,\frac{w}{w+1})$
\end{enumerate}
\end{tt}
\end{minipage}
}}
\end{center}

\begin{comment}
\begin{center}
\fbox{\parbox{5.3in}{
\begin{minipage}{5.3in}
\begin{tt}
{\bf Oracle}
\begin{enumerate}
\item[1)] Input: $(x_{1},...,x_{n},y)$
\item[2)] Output: A or B or mixed
\item[3)] If $\sum x_{i}>\frac{w}{w+1}n$ and $y>\frac{w}{w+1}$ return A.
\item[4)] If $\sum x_{i}<\frac{w}{w+1}n$ and $y<\frac{w}{w+1}$ return B.
\item[5)] Set $c_i = \frac{x_i(1-x_1)}{x_1(1-x_i)}$ for $i \geq 2$ and $c_1=1$.
\item[6)] Solve equation $\sum_{i=1}^n \frac{x_1'c_i}{1+(c_i-1)x_1'} = \frac{w}{w+1}n$ (binary search) \\ to compute $x_1'$ and set $x_i' = \frac{c_ix_1'}{1+(c_i-1)x_1'}$ for $i\geq 2$.
\item[7)] Let $f(t) = \left(\frac{t(w+1)}{w}\right)^{\frac{w}{w+1}}\left[(1-t)(w+1)\right]^{\frac{1}{w+1}}- \prod_{i} \left(\frac{x_{i}}{x_{i}'}\right)^{x_{i}'}\left(\frac{1-x_{i}}{1-x_{i}'}\right)^{1-x_{i}'}$.
\item[8)] If ($\sum_{i}x_{i}>\frac{w}{w+1}n$ and $f(y)<0$) or \\($\sum_{i}x_{i}<\frac{w}{w+1}n$ and $f(y)>0$) return B.
\item[9)] If ($\sum_{i}x_{i}>\frac{w}{w+1}n$ and $f(y)>0$) or \\($\sum_{i}x_{i}<\frac{w}{w+1}n$ and $f(y)<0$) return A.
\item[10)] return mixed.
\end{enumerate}
\end{tt}
\end{minipage}
}}
\end{center}
\end{comment}

The proof of correctness of the algorithm is presented in Appendix \ref{appendix:Star}.
Given this oracle, we establish an upper bound of $1.42$ for
 the average price of anarchy,  which is independent both of $w$ as well as the size of the star.

\begin{corollary}
\label{thm:1868}
 The average price of anarchy for the class of star $w$-coordination games (with $n+1$ agents) is at most  $1.42$.
% For any $w$, it converges to $1$  as $n$ goes to infinity.
 \end{corollary}
\begin{comment}
\begin{proof}
There are exactly two possible outcomes with positive probability; all the agents choose strategy $A$ and all choose strategy $B$. Assume we take one sample at random $(x_1,...,x_n,y)$ from $\times_{i=1}^{n+1}\Delta_2$ where $n+1$ are the number of players. It turns out from the oracle above on the star-graph (see also discussion in the full version)%appendix)
game that if $\sum_i x_i < n\frac{w}{w+1}$ and $y< \frac{w}{w+1}$ then the dynamics eventually converge to all agents choose $B$. Hence the region of attraction of the outcome all agents choose $B$ will be at least the probability that a sample at random satisfies $\sum_i x_i < n\frac{w}{w+1}$ and $y< \frac{w}{w+1}$. By Chernoff Bounds, this is at least $p = \frac{w}{w+1} (1 - e^{-n/3 \cdot (1/2- \frac{1}{w+1})})$. Since the optimal is $w(n+1)$, we get that the average price of anarchy  is at most $\frac{w(n+1)}{pw(n+1)+(1-p)(n+1)} = \frac{w}{pw + 1-p}$. It is not hard to see that $p$ is decreasing w.r.t $n,w$. For $n=1$ we have already established an upper bound on $1.21$, hence the maximum average price of anarchy is attained at $n=2$, for $w=2$ and it is $1.868$. Observe also that as $n\to \infty$, the average price of anarchy converges to $\frac{w^2+w}{w^2+1}$.
\end{proof}
\end{comment}
\subsection{Average price of anarchy of linear, symmetric load balancing games}
\label{sec:APoAload}

In this subsection, we state the following bounds on the average price of anarchy of linear, symmetric load balancing games.

\begin{theorem}
\label{thm:AvePoA}
The average price of anarchy in terms of makespan of symmetric, linear load balancing games is at most $3/2$.
Moreover, generically, the average price of anarchy of symmetric, linear load balancing games is $1$.
Specifically, given any number of agents and machines, the set of linear latency functions such that the average price of anarchy of the resulting game is greater than $1$ is a zero measure set within the set of all linear latency functions.
\end{theorem}

\noindent
For the classic game of $N$-balls $N$-bins we can show the following theorem:
\begin{theorem} The average price of anarchy in terms of makespan for the (identical) $N$-balls $N$-bins is $1$.
\label{thm:ballsapoa}
\end{theorem}

\section{Conclusion and Open questions}\label{sec:conclusion}

We define an average case analysis notion in dynamical systems focusing on games and replicator dynamics. We call this notion average price of anarchy (APoA) and provide upper and lower bounds for APoA in different classes of games. Several questions arise:
% Along these lines, the dynamics we focus on is called replicator dynamics. This dynamics is inspired from evolution and has some nice properties (stable fixed points are Nash equilibria, convergence). There are many questions that someone could look at, but we mention the ones that we believe are very interesting and important.
\begin{itemize}
\item \textbf{Other settings/games/mechanisms.} In recent followup work, \cite{Shnayder_ijcai16} applies our approach to peer prediction mechanisms where the size of the basin of attraction of the truthful equilibrium is used as a proxy for the robustness of truthful play. The replicator model predicts/confirms the significant improvement in robustness of recent mechanisms over earlier approaches. It would be interesting to test the robustness of other (approximately) truthful, differentially private mechanisms in a similar manner.

\item \textbf{Other dynamics.} %Replicator dynamics has some nice properties and all the machinery of this paper works for it.
Perform average case analysis for other dynamics and compare them against replicator dynamics.
\item \textbf{Generalization of APoA.} Generalize the notion of APoA to dynamics that do not necessarily converge.  In particular, it would be intriguing to define an APoA notion for chain recurrent sets (see \cite{Papadimitriou:2016:NEC:2840728.2840757}).
\item \textbf{Point-wise convergence.} Generalize the point-wise convergence result to a larger class of congestion games, (e.g., for polynomial cost functions), as well as  extend the point-wise convergence result for linear cost functions to other dynamics.
\item \textbf{Volumes of regions of attraction as a function.} Given a prior distribution over initial conditions (e.g., uniform), every point-wise convergent dynamical system induces a probability distribution over fixed points. By approximating this function (from priors over initial conditions to posteriors over equilibria), we can predict the average case (long-term) behavior of the system (without having the equations of the dynamics). This interpretation of a game, as an experiment/measurement that maps the outside observer's original beliefs over initial mixed strategies/beliefs of the agents to (sampling from) a distribution/belief over the resulting equilibria,  resolves the non-determinism problem linked to the multiplicity  equilibria in a game. The unique, well-defined, deterministic prediction is the function from distributions over initial conditions to distributions over equilibria. Naturally, if the initial distribution is concentrated on a single equilibrium then the output of the prediction function will similarly be concentrated on that equilibrium. Nontrivial distributions will result to a (unique) distribution/prediction that puts positive measure on several equilibria. Developing a formal theory for approximating such functions seems like a fertile ground for combining ideas from computer science and game theory.
\end{itemize}

\section*{Acknowledgments}
The authors would like to thank Iosif Pinelis for the proof of Claim \ref{cl:overflow}. 
\bigskip

 Ioannis Panageas was supported by NSF EAGER award grants CCF-1415496 and CCF-1415498. Georgios Piliouras would like to acknowledge
 %AFOSR/MURI Project FA9550-09-1-0538, AFOSR Project FA9550-09-1-0420 as well as  ONR Grant N00014-09-1-0751,
 the CMI Wally Baer and Jeri Weiss postdoctoral fellowship, SUTD grant SRG ESD 2015 097 and MOE AcRF Tier 2 Grant  2016-T2-1-170. Part of the work was completed while Georgios Piliouras was a CMI Wally Baer and Jeri Weiss postdoctoral fellow at California Institute of Technology.
 Part of the work was completed while Ioannis Panageas and Georgios Piliouras were resident scientists at the Simons Institute for 
 the Theory of Computing.

% Bibliography
\bibliographystyle{plain}
\bibliography{sigproc2ecb}

%\newpage

\section*{APPENDIX}

%\appendix
%\setcounter{section}{1}

\section{Missing proofs and lemmas from Sections \ref{sec:definitions} and \ref{sec:tools}}

\begin{lemma}\label{lem:measurable} $\psi(\vec{x})$ is a measurable function.
\end{lemma}

\begin{proof} For an arbitrary $c \in \mathbb{R}$ we have that $$\{\vec{x}: \psi(\vec{x})_{i}<c\} = \cup_{k=1}^{\infty}\cup_{m=1}^{\infty}\cap_{n>m}^{\infty}\{\vec{x}: \phi_n(\vec{x})_{i}< c -\frac{1}{k}\}\}.$$
The set $\{\vec{x}: \phi_n(\vec{x})_{i}< c -\frac{1}{k}\}$ is measurable since $\phi_n(\vec{x})_{i}$ is a (Lebesgue) measurable function (by continuity). Therefore $\psi(\vec{x})_{i}$ is a measurable function.
\end{proof}

\subsection{Proof of Theorem \ref{thm:congestionpointwise} for network coordination games}
\label{appendix:congestionpointwise}

\begin{proof}
We denote by $\hat{u}_{i}$ the expected utility of agent $i$ under mixed strategy profile $\vec{p}$ and by $u_{i\gamma}$ his expected utility when he deviates to strategy $\gamma$ and all other agents still play according to $\vec{p}$.
We observe that
\begin{align*}
\Psi (\vec{p}) &= \sum_{i}\hat{u}_{i}=\sum_{i,\gamma}p_{i\gamma}\sum_{j \in N(i)}\sum_{\delta}A_{ij}^{\gamma\delta}p_{j\delta}
\end{align*} is a Lyapunov function for our game since (strictly increasing along the trajectories)
\begin{align*}
\frac{\partial \Psi}{\partial p_{i\gamma}} &=  u_{i\gamma} + \sum_{j \in N(i)}\sum_{\delta}A_{ji}^{\delta\gamma}p_{j\delta}=2u_{i\gamma}\;\textrm{ since }A_{ij} = A_{ji}^T
\end{align*}
and hence
\begin{align*}
\frac{d \Psi}{dt} &= \sum_{i,\gamma}\frac{\partial \Psi}{\partial p_{i\gamma}}\frac{d p_{i\gamma}}{dt} = \sum_{i,\gamma,\gamma'}p_{i\gamma}p_{i\gamma'}(u_{i\gamma}-u_{i\gamma'})^2 \geq 0
\end{align*}
 with equality at fixed points. Hence (as in \cite{Kleinberg09multiplicativeupdates}) we have convergence to equilibrium sets (compact connected sets consisting entirely of fixed points). We address the fact that this doesn't suffice for pointwise convergence. To be exact it suffices only in the case the equilibria are isolated (which is not the case for network coordination games - see figure \ref{fig:gamestar}).\\

Let $\vec{q}$ be a limit point of the trajectory $\vec{p} (t)$.
W.l.o.g we can assume that $\vec{p} (0)$ is in the interior of $\Delta$  and hence $\vec{p} (t)$ is in the interior of $\Delta$ for all $t \in \mathbb{R}$  (we can assume that we start in the interior of $\Delta$ otherwise we can just consider the subgame defined by the strategies that agents play with positive probability.).
We have that $\Psi(\vec{q})\geq\Psi(\vec{p} (t))$ where the equality holds only if we start at equilibrium.
 We define the relative entropy. $$I(\vec{p}) = -\sum_{i}\sum_{\gamma : q_{i\gamma}>0}q_{i\gamma} \ln (p_{i\gamma}/q_{i\gamma}) \geq 0 \textrm{ (Jensen's inequality)}$$
and $I(\vec{p})=0$ iff $\vec{p} = \vec{q}$. We get that
\begin{align*}
\frac{d I}{dt} &= -\sum_{i}\sum_{\gamma : q_{i\gamma}>0}q_{i\gamma}(u_{i\gamma}-\hat{u}_{i}) \\
%&= \sum_{i} \hat{u}_{i} - \sum_{i}\sum_{\gamma: q_{i\gamma}>0}q_{i\gamma}u_{i\gamma}\\
&= \sum_{i} \hat{u}_{i} - \sum_{i,\gamma}q_{i\gamma}u_{i\gamma}
%\\&=
%-\sum_{i} \hat{c}_{i} + \sum_{i}\sum_{\gamma}q_{i\gamma}[\sum_{j \neq i}\sum_{\gamma'}\sum_{e \in \gamma \cap \gamma %'}a_{e}p_{j\gamma '}+\sum_{e\in \gamma}(b_{e}+a_{e})]\\
\\&=
\sum_{i} \hat{u}_{i} - \sum_{i,\gamma}\sum_{j \in N(i)}\sum_{\delta}A_{ij}^{\gamma\delta}p_{j\delta}q_{i\gamma}
\\&=
\sum_{i} \hat{u}_{i} - \sum_{j,\delta}\sum_{i \in N(j)}\sum_{\gamma}A_{ij}^{\gamma\delta}p_{j\delta}q_{i\gamma} \;\textrm{ (since }A_{ij} = A_{ji}^T)
%\\&=
%-\sum_{i} \hat{c}_{i} + \sum_{i}\sum_{\gamma}\sum_{e\in \gamma}(b_{e}+a_{e})q_{i\gamma} + %\sum_{i}\sum_{\gamma}p_{i\gamma}(d_{i\gamma} - \sum_{e\in \gamma}b_{e}+a_{e})
\\&=
\sum_{i} \hat{u}_{i} - \sum_{j,\delta}p_{j\delta}d_{j\delta}\\&=
\sum_{i}\hat{u}_{i}-\sum_{i} \hat{d}_{i}  - \sum_{j,\delta}p_{j\delta}(d_{j\delta}-\hat{d}_{j})
%\\&=\sum_{i}\hat{d}_{i} + \sum_{i}\sum_{\gamma}\sum_{e\in \gamma}(b_{e}+a_{e})q_{i\gamma} - \sum_{i} \hat{c}_{i} - %\sum_{i}\sum_{\gamma}\sum_{e\in \gamma}(b_{e}+a_{e})p_{i\gamma}- \sum_{i}\sum_{\gamma}p_{i\gamma}(\hat{d}_{i}-d_{i\gamma})
\\&= \Psi(\vec{p}) - \Psi(\vec{q})- \sum_{i,\gamma}p_{i\gamma}(d_{i\gamma}-\hat{d}_{i})
\end{align*}
where $d_{i\gamma}, \hat{d}_{i}$ correspond to the payoff of agent $i$ if he chooses strategy $\gamma$ and his expected payoff respectively at point $\vec{q}$. The rest of the proof follows in a similar way to Losert and Akin \cite{akin}.\\

We break the term $\sum_{i,\gamma}p_{i\gamma}(d_{i\gamma}-\hat{d}_{i})$ to positive and negative terms (we ignore zero terms), i.e., $\sum_{i,\gamma}p_{i\gamma}(d_{i\gamma}-\hat{d}_{i}) = \sum_{i,\gamma: \hat{d}_{i} >d_{i\gamma}}p_{i\gamma}(d_{i\gamma}-\hat{d}_{i}) + \sum_{i,\gamma: \hat{d}_{i} <d_{i\gamma}}p_{i\gamma}(d_{i\gamma}-\hat{d}_{i})$.
\\\\ \textbf{Claim:} There exists an $\epsilon >0$ so that the function $Z(\vec{p}) = I(\vec{p})+ 2\sum_{i,\gamma: \hat{d}_{i}>d_{i\gamma}}p_{i,\gamma}$ has $\frac{dZ}{dt}<0$ for $\norm[1]{\vec{p}-\vec{q}}<\epsilon$ and $\Psi(\vec{q})>\Psi(\vec{p})$.\\\\ Assuming that $\vec{p} \to \vec{q}$, we get $u_{i\gamma}-\hat{u}_{i}  \to d_{i\gamma}-\hat{d}_{i}$ for all $i,\gamma$. Hence for small enough $\epsilon>0$ with $\norm[1]{\vec{p}-\vec{q}}<\epsilon$, we have that $u_{i\gamma}-\hat{u}_{i} \leq \frac{3}{4} (d_{i\gamma}-\hat{d}_{i})$ for the terms which $d_{i\gamma}-\hat{d}_{i}<0$. Therefore
\begin{align*}
\frac{dZ}{dt} &=  \Psi(\vec{p}) - \Psi(\vec{q}) - \sum_{i,\gamma: \hat{d}_{i} <d_{i\gamma}}p_{i\gamma}(d_{i\gamma}-\hat{d}_{i}) - \sum_{i,\gamma: \hat{d}_{i} >d_{i\gamma}}p_{i\gamma}(d_{i\gamma}-\hat{d}_{i}) + 2\sum_{i,\gamma: \hat{d}_{i}>d_{i\gamma}}p_{i\gamma}(u_{i\gamma}-\hat{u}_{i}) \\&\leq \Psi(\vec{p}) - \Psi(\vec{q}) - \sum_{i,\gamma: \hat{d}_{i} <d_{i\gamma}}p_{i\gamma}(d_{i\gamma}-\hat{d}_{i}) - \sum_{i,\gamma: \hat{d}_{i} >d_{i\gamma}}p_{i\gamma}(d_{i\gamma}-\hat{d}_{i}) + 3/2\sum_{i,\gamma: \hat{d}_{i} >d_{i\gamma}}p_{i\gamma}(d_{i\gamma}-\hat{d}_{i}) \\&=
\underbrace{\Psi(\vec{p}) - \Psi(\vec{q})}_{< 0} + \underbrace{\sum_{i,\gamma: \hat{d}_{i} <d_{i\gamma}}-p_{i\gamma}(d_{i\gamma}-\hat{d}_{i})}_{\leq 0} + 1/2\underbrace{\sum_{i,\gamma: \hat{d}_{i} >d_{i\gamma}}p_{i\gamma}(d_{i\gamma}-\hat{d}_{i})}_{\leq 0}<0
\end{align*}
where we substitute $\frac{p_{i\gamma}}{dt} = p_{i\gamma}(u_{i\gamma}-\hat{u}_{i})$ (replicator), and the claim is proved. \\\\
Note that $Z(\vec{p}) \geq 0$ (sum of nonnegative terms and $I(\vec{p}) \geq 0$) and is zero iff $\vec{p} = \vec{q}.\;$ (i)

To finish the proof of the theorem, if $\vec{q}$ is a limit point of $\vec{p} (t)$, there exists an increasing sequence of times $t_{n}$, with $t_{n} \to \infty$ and $\vec{p} (t_{n}) \to \vec{q}$. We consider $\epsilon '$ such that the set $C = \{\vec{p} : Z(\vec{p})< \epsilon '\}$ is inside $B = \norm[1]{\vec{p}-\vec{q}}<\epsilon$ where $\epsilon$ is from claim above. Since $\vec{p}( t_{n}) \to \vec{q}$, consider a time $t_{N}$ where $\vec{p} (t_{N})$ is inside $C$. From claim above we get that $Z(\vec{p})$ is decreasing inside $B$ (and hence inside $C$), thus $Z(\vec{p} (t)) \leq Z(\vec{p} (t_{N})) < \epsilon '$ for all $t \geq t_{N}$, hence the orbit will remain in $C$. By the fact that $Z(\vec{p} (t))$ is decreasing in $C$ (claim above) and also $Z(\vec{p} (t_{n})) \to Z(\vec{q}) =0$ it follows that $Z(\vec{p} (t)) \to 0$ as $t \to \infty$. Hence $\vec{p} (t) \to \vec{q}$ as $t \to \infty$ using (i).
\end{proof}

\subsection{Proof of Theorem \ref{zero}}
\label{appendix:zero}

To prove the theorem we will use the Center-Stable Manifold theorem (see Theorem \ref{manifold}). In order to do that we need a map whose domain is full-dimensional. However, a simplex in $\mathbb{R}^n$ has dimension $n-1$. Therefore, we need to take a projection of the domain space and accordingly redefine the map of the dynamical system. We note that the projection we take will be fixed point dependent; this is to keep the proof that every stable fixed point is a weakly stable Nash proved in \cite{Kleinberg09multiplicativeupdates} relatively less involved later. Let $\vec{q}$ be a point of our state space $\Delta$ and $\Sigma = |\cup_{i}S_{i}|$. Let $h_{\vec{q}} : [N] \to [\Sigma]$ be a function such that $h_{\vec{q}}(i) = \gamma$ if $q_{i\gamma}> 0$ for some $\gamma \in S_{i}$. Let $M = \sum |S_{i}|$ and $g$ a fixed projection where you exclude the first coordinate of every agent's distribution vector. We consider the mapping $z_{\vec{q}} : R^M \to R^{M-N}$ so that we exclude from each agent $i$ the variable $p_{i,h_{\vec{q}}(i)}$ ($z_{\vec{q}}$ plays the same role as $g$ but we drop variables with a specific property this time, those played with positive probability). We substitute the variables $p_{i, h_{\vec{q}}(i)}$ with  $1 - \sum_{\gamma \in S_{i} \atop \gamma \neq h_{\vec{q}}(i)}p_{i \gamma}$.

\noindent
The  formal statement of the Center-Stable manifold theorem  %which is of great importance in dynamical systems
has as follows:

\begin{theorem}\label{manifold}(Center and Stable Manifolds, p. 65 of \cite{shub})
Let $\vec{0}$ be a fixed point for the $C^r$ local differomorphism $f: U \to \mathbb{R}^n$ where $U \subset \mathbb{R}^n$ is a neighborhood of zero in $\mathbb{R}^n$ and $r \geq 1$. Let $E^s \oplus E^c \oplus E^u$ be the invariant splitting of $\mathbb{R}^n$ into generalized eigenspaces of $Df(\vec{0})$ corresponding to eigenvalues of absolute value less than one, equal to one, and greater than one. To the $Df(\vec{0})$ invariant subspace $E^s\oplus E^c$ there is associated a local $f$ invariant $C^r$ embedded disc $W^{sc}_{loc}$ tangent to the linear subspace at $\vec{0}$ and a ball $B$ around zero such that:
\begin{equation}
f(W^{sc}_{loc}) \cap B \subset W^{sc}_{loc}.\textrm{  If } f^n(\vec{x}) \in B \textrm{ for all }n \geq 0, \textrm{ then }\vec{x} \in W^{sc}_{loc}
\end{equation}
\end{theorem}
For $t=1$ and an unstable fixed point $\vec{p}$ we consider the function $\psi_{1,\vec{p}}(\vec{x}) = z_{\vec{p}}\circ \phi_{1} \circ z_{\vec{p}}^{-1}(\vec{x})$ which is $C^1$ diffeomorphism, where $\phi_1$ is the time one map of the flow of the dynamical system in $\Delta$ (we assume we do the renormalization trick described in Section \ref{sec:systems}).
%, and $U \subset \mathbb{R}^{M-N}$ an open ball around $z_{\vec{p}}(\vec{p})$ that contains $z_{\vec{p}}(\Delta)$ (it will contain points with negative sign as well).
Let $B_{z_{\vec{p}}(\vec{p})}$ be the ball that is derived from \ref{manifold} and we consider the union of these balls (transformed in $\mathbb{R}^M$) $$A = \cup _{\vec{p}}A_{z_{\vec{p}}(\vec{p})}$$ where $A_{z_{\vec{p}}(\vec{p})} = g \circ z_{\vec{p}}^{-1}(B_{z_{\vec{p}}(\vec{p})})$ ($z^{-1}_{\vec{p}}$ "returns" the set $B_{z_{\vec{p}}(\vec{p})}$ back to $\mathbb{R}^M$).
Taking advantage of separability of $\mathcal{R}^M$ we have the following theorem.
\begin{theorem} (Lindel\H{o}f's lemma) For every open cover there is a countable subcover.
\end{theorem}
Therefore we can find a countable subcover for $A = \cup _{\vec{p}}A_{z_{\vec{p}}(\vec{p})}$, i.e., $A = \cup _{m=1}^{\infty}A_{z_{\vec{p}_{m}}(\vec{p}_{m})}$.
\\\\ Let $\psi_{n,\vec{p}}(\vec{x}) = z_{\vec{p}}\circ \phi_{n} \circ z_{\vec{p}}^{-1}(\vec{x})$. If a point $\vec{x} \in  g(\Delta)$ (which corresponds to $g^{-1}(\vec{x})$ in our original $\Delta$) has as unstable fixed point as a limit, there must exist a $n_{0}$ and $m$ so that $\psi_{n,\vec{p}_{m}} \circ z_{\vec{p}_{m}} \circ g^{-1}(\vec{x}) \in B_{ z_{\vec{p}_{m}}(\vec{p}_{m})}$ for all $n \geq n_{0}$ and therefore again from \ref{manifold} and the fact that $\Delta$ is invariant we get that we get that $\psi_{n_{0},\vec{p}_{m}} \circ z_{\vec{p}_{m}} \circ g^{-1}(\vec{x}) \in (W_{loc\;z_{\vec{p}_{m}}(\vec{p}_{m})}^{sc} \cap z_{\vec{p}_{m}}(\Delta))$, hence $\vec{x} \in g\circ z^{-1}_{\vec{p}_{m}} \circ \psi^{-1}_{n_{0},\vec{p}_{m}}(W_{loc\;z_{\vec{p}_{m}}(\vec{p}_{m})}^{sc} \cap z_{\vec{p}_{m}}(\Delta))$.\\\\
Hence, the set of points in $g(\Delta)$ whose $\omega$-limit has an unstable equilibrium, is a subset of
\begin{equation} C=  \cup_{m=1}^{\infty} \cup_{n=1}^{\infty} g\circ z^{-1}_{\vec{p}_{m}} \circ \psi^{-1}_{n,\vec{p}_{m}}(W_{loc\;z_{\vec{p}_{m}}(\vec{p}_{m})}^{sc}\cap z_{\vec{p}_{m}}(\Delta))
\end{equation}
Observe that the dimension of $W_{loc\;z_{\vec{p}_{m}}(\vec{p}_{m})}^{sc}$ is at most $M-N-1$ since we assume that $\vec{p}_{m}$ is unstable ($\mathcal{J}_{\vec{p}_{m}}$ has an eigenvalue with positive real part)\footnote{Here we used the fact that the eigenvalues with absolute value less than one, one and greater than one of $e^A$ correspond to eigenvalues with negative real part, zero real part and positive real part respectively of $A$} and thus $dim E^u \geq 1$, hence the set $(W_{loc\;z_{\vec{p}_{m}}(\vec{p}_{m})}^{sc}\cap z_{\vec{p}_{m}}(\Delta))$  has Lebesgue measure zero in $\mathbb{R}^{M-N}$. Finally since $g\circ z^{-1}_{\vec{p}_{m}} \circ \psi^{-1}_{n,\vec{p}_{m}} : \mathbb{R}^{M-N} \to \mathbb{R}^{M-N})$ is continuously differentiable in an open neighborhood of $g(\Delta)$, $\psi_{n, \vec{p}_{m}}$ is $C^1$ and hence
locally Lipschitz in that neighborhood (see \cite{perko} p.71) and it preserves the null-sets (see Lemma \ref{lips}). Namely, $C$ is a countable union of measure zero sets, i.e., is measure zero as well. Since the dynamical system after renormalization is topologically equivalent with the system before renormalization, Theorem \ref{zero} follows.
\begin{lemma}\label{lips} Let $g: \mathbb{R}^n \to \mathbb{R}^n$ be a locally  Lipschitz function, then $g$ is null-set preserving, i.e., for $E \subset \mathbb{R}^n$ if $E$ has measure zero then $g(E)$ has also measure zero.
\end{lemma}
\begin{proof}  Let $B_{\gamma}$ be an open ball such that $\norm[2]{g(\vec{y}) - g(\vec{x})} \leq K_{\gamma} \norm[2]{\vec{y}-\vec{x}}$ for all $\vec{x},\vec{y} \in B_{\gamma}$. We consider the union $\cup_{\gamma}B_{\gamma}$ which cover $\mathbb{R}^n$ by the assumption that $g$ is locally Lipschitz. By Lindel\H{o}f's lemma we have a countable subcover, i.e., $\cup_{i=1}^{\infty}B_{i}$. Let $E_{i} = E \cap B_{i}$. We will prove that $g(E_{i})$ has measure zero. Fix an $\epsilon >0$. Since $E_{i} \subset E$, we have that $E_{i}$ has measure zero, hence we can find a countable cover of open balls $C_{1},C_{2},\dots $ for $E_{i}$, namely $E_{i} \subset \cup_{j=1}^{\infty}C_{j}$ so that $C_{j} \subset B_{i}$ for all $j$ and also $\sum_{j=1}^{\infty} \mu(C_{j}) < \frac{\epsilon}{K_{i}^n}$. Since $E_{i} \subset \cup_{j=1}^{\infty}C_{j}$ we get that $g(E_{i}) \subset \cup_{j=1}^{\infty}g(C_{j})$, namely $g(C_{1}),g(C_{2}),\dots$ cover $g(E_{i})$ and also $g(C_{j}) \subset g(B_{i})$ for all $j$. Assuming that ball $C_{j} \equiv B(\vec{x},r)$ (center $\vec{x}$ and radius $r$) then it is clear that $g(C_{j}) \subset B(g(\vec{x}),K_{i} r)$ ($g$ maps the center $\vec{x}$ to $g(\vec{x})$ and the radius $r$ to $K_{i}r$ because of Lipschitz assumption). But $\mu(B(g(\vec{x}),K_{i} r)) = K_{i}^n \mu(B(\vec{x}, r)) = K_{i}^n \mu(C_{j})$, therefore $\mu(g(C_{j})) \leq K_{i}^n \mu(C_{j})$ and so we conclude that $$\mu(g(E_{i})) \leq \sum_{j=1}^{\infty}\mu(g(C_{j})) \leq K_{i}^n \sum_{j=1}^{\infty}\mu(C_{j}) < \epsilon$$
Since $\epsilon$ was arbitrary, it follows that $\mu(g(E_{i})) =0$. To finish the proof, observe that $g(E) = \cup_{i=1}^{\infty} g(E_{i})$ and therefore $\mu(g(E)) \leq \sum_{i=1}^{\infty} \mu(g(E_{i})) =0$.
\end{proof}

\subsection{Proof of Lemma \ref{lemma:KL}}
\label{appendix:KL}

\begin{proof}
The derivative of  $\sum_{i\in V_{left}}\sum_{\gamma \in S_i} q_{i\gamma} \cdot \ln(p_{i\gamma})-\sum_{i\in V_{right}}\sum_{\gamma \in S_i} q_{i\gamma} \cdot \ln(p_{i\gamma})$ has as follows:

\begin{eqnarray*}
\lefteqn{\sum_{i \in V_{left}}  \sum_{\gamma \in S_i} q_{i\gamma}  \frac{d\ln(p_{i\gamma})}{dt} - \sum_{i \in V_{right}}  \sum_{\gamma \in S_i} q_{i\gamma}  \frac{d\ln(p_{i\gamma})}{dt}= \sum_{i \in V_{left}}  \sum_{\gamma \in S_i}q_{i\gamma}  \frac{\dot{p}_{i\gamma}}{p_{i\gamma}}  -  \sum_{i \in V_{right}}  \sum_{\gamma\in S_i}q_{i\gamma}  \frac{\dot{p}_{i\gamma}}{p_{i\gamma}}= }\\
%&=&  \sum_{i \in V_{left}}  \sum_{\gamma\in S_i}q_{i\gamma}  \frac{\dot{p}_{i\gamma}}{p_{i\gamma}}  -  \sum_{i \in V_{right}}  \sum_{\gamma \in S}q_{i\gamma}  \frac{\dot{p}_{i\gamma}}{p_{i\gamma}} =\\
%\sum_i  a_i \big(\sum_{R \in S_i} q_{iR} u^i(R) - \sum_{R \in S_i} p_{i\gamma}u^i(R) \big)=}\\ %[u_i(x) - \hat{u}(x)]}\\  %sum_{(i,k)\in E} \big(q^\mathrm{T}_i A^{i,k} x_k -x^\mathrm{T}_i A^{i,k} x_k \big)=}\\
 %&=& \sum_i \big(\sum_{R \in S_i} q_{iR}(a_i  u^i(R)+b_i) - \sum_{R \in S_i} x_{iR}(a_i u^i(R)+b_i) \big) =\\
 &=&\sum_{i \in V_{left}}  \sum_{(i,j)\in E} \big(\vec{q_i}^\mathrm{T} A_{ij} \vec{p_j} -\vec{p_i}^\mathrm{T} A_{ij} \vec{p_j} \big) - \sum_{i \in V_{right}}  \sum_{(i,j)\in E} \big(\vec{q_i}^\mathrm{T}A_{ij} \vec{p_j} -\vec{p_i}^\mathrm{T} A_{ij} \vec{p_j} \big)=\\%=\\
 &=& \sum_{i \in V_{left}}  \sum_{(i,j)\in E} \big(\vec{q_i}^\mathrm{T} -\vec{p_i}^\mathrm{T}\big)A_{ij} \vec{p_j} - \sum_{i \in V_{right}}  \sum_{(i,j)\in E} \big(\vec{q_i}^\mathrm{T} -\vec{p_i}^\mathrm{T}\big)A_{ij} \vec{p_j} =\\  %=\\
 &=&  \sum_{i \in V_{left}} \sum_{(i,j)\in E} \big(\vec{q_i}^\mathrm{T} -\vec{p_i}^\mathrm{T}\big)A_{ij} (\vec{p_j}-\vec{q_j}) - \sum_{i \in V_{right}} \sum_{(i,j)\in E} \big(\vec{q_i}^\mathrm{T} -\vec{p_i}^\mathrm{T}\big)A_{ij} (\vec{p_j}-\vec{q_j})=\\
 %&=& - \sum_i \sum_{(i,k)\in E} \big(q^\mathrm{T}_i -x^\mathrm{T}_i\big)A^{i,k} (q_k-x_k)=\\
% &\geq&  \sum_i \sum_{(i,k)\in E} \big(q^\mathrm{T}_i -x^\mathrm{T}_i\big)A^{i,k} (x_k-q_k)\\
%&=& - \sum_{(i,k)\in E, i<k}\big[ \big(q^\mathrm{T}_i -x^\mathrm{T}_i\big)A^{i,k} (q_k-x_k) +
%\big(q^\mathrm{T}_k -x^\mathrm{T}_k\big)A^{k,i} (q_i-x_i)\big] =\\
%&=& - \sum_{(i,k)\in E, i<k}\big[ \big(q^\mathrm{T}_i -x^\mathrm{T}_i\big)A^{i,k} (q_k-x_k) +
%\big(q^\mathrm{T}_i -x^\mathrm{T}_i\big)\big(A^{k,i}\big)^\mathrm{T} (q_k-x_k) \big] =\\
&=& - \sum_{(i,j)\in E, i\in V_{left}, j\in V_{right}}\big[ \big(\vec{q_i}^\mathrm{T} -\vec{p_i}^\mathrm{T}\big)A_{ij} (\vec{q_j}-\vec{p_j}) -
\big(\vec{q_j}^\mathrm{T} -\vec{p_j}^\mathrm{T}\big) %\big(
A_{ji}%\big)
(\vec{q_i}-\vec{p_i}) \big] =
 0\\
%&=& 0
\end{eqnarray*}

\noindent
where the second to last line follows from the fact that $\vec{q} = (\vec{q}_{1},\dots,\vec{q}_{N})$ is a fully mixed Nash equilibrium. The last equality follows from the fact that all edge/games are coordination/partnership games, i.e., $A_{ji}^\mathrm{T}=A_{ij}$.
\end{proof}

\section{Missing proofs and lemmas of section~\ref{sec:average}}

%\section{Missing proofs of section \ref{sec:stability}}

%\section{Missing proofs of section \ref{sec:invariants}}

\subsection{Missing proof of technical lemma in \ref{sec:staghunt}}

%We establish that the curve
% $p_{2s}=\frac12 (1 - p_{1s} + \sqrt{1 + 2 p_{1s} - 3 p_{1s}^2})$ captures the stable manifold of the mixed Nash equilibrium.
 %It is trivial to check that the mixed Nash satisfies this equation. Furthermore,

 The following technical lemma argues that the vector filed of the replicator dynamic in the Stag Hunt game is  tangent to the
 curve
 $p_{2s}=\frac12 (1 - p_{1s} + \sqrt{1 + 2 p_{1s} - 3 p_{1s}^2})$.
 % and hence by uniqueness of the solution of the replicator ODE captures the stable manifold of the mixed Nash.

 \begin{lemma}
 \label{lemma:tangent}
 For any $0<p_{1s}, p_{2s}<1$, with $p_{2s}=\frac12 (1 - p_{1s} + \sqrt{1 + 2 p_{1s} - 3 p_{1s}^2})$   we have that:
 $$\frac{\partial p_{2s}}{\partial p_{1s}}=  \frac{\frac{dp_{2s}}{dt}}{\frac{dp_{1s}}{dt}}=\frac{p_{2s}\big(u_2(s)-(p_{2s}u_2(s)+(1-p_{2s})u_2(h)) \big)}{p_{1s}\big(u_1(s)-(p_{1s}u_1(s)+(1-p_{1s})u_1(h))\big)}$$
 %
 %$$\frac{\partial p_{2s}}{\partial p_{1s}}= \frac{\zeta_{2h}}{\zeta_{1h}}=\frac{p_{2s}\big(u_2(h)-(p_{2s}u_2(h)+(1-p_2(h))u_2(s)) \big)}{p_{1s}\big(u_1(h)-(p_{1s}u_1(h)+(1-p_1(h))u_1(s))\big)}.$$
 \end{lemma}

\begin{proof}
By substitution of the Stag Hunt game utilities, we have that:

\begin{equation}
\label{eq:tan1}
 %\frac{\zeta_{2h}}{\zeta_{1h}}=\frac{p_{2s}\big(u_2(h)-(p_{2s}u_2(h)+(1-p_2(h))u_2(s)) \big)}{p_{1s}\big(u_1(h)-(p_{1s}u_1(h)+(1-p_1(h))u_1(s))\big)}
 %\frac{\partial p_{2s}}{\partial p_{1s}}=
 \frac{\zeta_{2s}}{\zeta_{1s}}=\frac{p_{2s}\big(u_2(s)-(p_{2s}u_2(s)+(1-p_{2s})u_2(h)) \big)}{p_{1s}\big(u_1(s)-(p_{1s}u_1(s)+(1-p_{1s})u_1(h))\big)}
 = \frac{p_{2s}(1-p_{2s})(3p_{1s}-2)}{p_{1s}(1-p_{1s})(3p_{2s}-2)}
\end{equation}

However, $p_{2s}(1-p_{2s})= \frac12 p_{1s}(p_{1s}-1 +\sqrt{1+2p_{1s}-3p^2_{1s}})$. Combining this with~(\ref{eq:tan1}),

\begin{equation}
\label{eq:tan2}
 \frac{\zeta_{2s}}{\zeta_{1s}}=\frac12 \frac{(p_{1s}-1+\sqrt{1+2p_{1s}-3p^2_{1s}})(3p_{1s}-2)}{(1-p_{1s})(3p_{2s}-2)}=\frac12 \frac{(\sqrt{1+3p_{1s}}-\sqrt{1-p_{1s}})(3p_{1s}-2)}{\sqrt{1-p_{1s}}\cdot(3p_{2s}-2)}
\end{equation}

Similarly,  we have that $3p_{2s}-2 = \frac12 \sqrt{1+3p_{1s}}\cdot (3\sqrt{1-p_{1s}}-\sqrt{1+3p_{1s}})$. By multiplying and dividing equation~(\ref{eq:tan2}) with $(\sqrt{1+3p_{1s}}+3\sqrt{1-p_{1s}})$ we get:

\begin{eqnarray}
\label{eq:tan3}
\nonumber
 \frac{\zeta_{2s}}{\zeta_{1s}}&=&\frac12 \frac{(\sqrt{1+3p_{1s}}+3\sqrt{1-p_{1s}})(\sqrt{1+3p_{1s}}-\sqrt{1-p_{1s}})(3p_{1s}-2)}{2\sqrt{1-p_{1s}}\cdot\sqrt{1+3p_{1s}}\cdot(2-3p_{1s})}\\
\nonumber
 &=& -\frac14  \frac{(\sqrt{1+3p_{1s}}+3\sqrt{1-p_{1s}})(\sqrt{1+3p_{1s}}-\sqrt{1-p_{1s}})}{\sqrt{1 + 2 p_{1s} - 3 p_{1s}^2})}\\ \nonumber
 &=& \frac12 \big(-1 + \frac{1-3p_{1s}}{\sqrt{1 + 2 p_{1s} - 3 p_{1s}^2}}\big)
 =  \frac{\partial \Big(\frac12 (1 - p_{1s} + \sqrt{1 + 2 p_{1s} - 3 p_{1s}^2})\Big)}{\partial p_{1s}}=\frac{\partial p_{2s}}{\partial p_{1s}}.
 \end{eqnarray}
\end{proof}

%\subsection{Proof of Theorem~\ref{thm:SH}}
%\begin{proof} See appendix.
%\end{proof}

\subsection{Proof of Theorem~\ref{thm:GenSH}}
\label{appendix:GenSH}

For any $w$, a $w$-coordination game is a potential game and therefore it is payoff equivalent to a congestion game. The only two weakly stable equilibria
are the pure ones, hence in order to understand the average case system performance it suffices to understand the size of regions of attraction for each of them.
As in the case of Stag Hunt game, we focus on the projection of the system to the subspace $(p_{1s},p_{2s})\subset [0,1]^2$.

We denote by $\zeta, \psi$, the projected flow and vector field respectively.

\begin{lemma}
All but a zero measure of initial conditions in the polytope $(P_{Hare})$:

%$$p_{2s}< -wp_{1s}+w \cap p_{2s} < -\frac{1}{w}p_{1s} +1$$

\begin{eqnarray*}
p_{2s}&\leq& -w p_{1s}+w\\
 p_{2s}&\leq& -\frac{1}{w}p_{1s} +1\\
 0&\leq& p_{1s},p_{2s} \leq1
 \end{eqnarray*}

\noindent
converges to the $(Hare, Hare)$ equilibrium. All but a zero measure of initial conditions in the polytope  $(P_{Stag})$:

\begin{eqnarray*}
p_{2s}&\geq& -p_{1s} +\frac{2w}{w+1} \\
%p_{2s}&<& -wp_{1s}+w\\
 %p_{2s}&<& -\frac{1}{w}p_{1s} +1\\
 0&\leq& p_{1s},p_{2s} \leq1
 \end{eqnarray*}

\noindent
converges to the $(Stag, Stag)$ equilibrium.
\end{lemma}

\begin{proof}
First, we will prove the claimed property for polytope $(P_{Stag})$. Since the game is symmetric, the replicator dynamics are similarly symmetric with $p_{2s}=p_{1s}$ axis of symmetry. Therefore it suffices to prove the property for the polytope $P'_{Hare} = P_{Hare} \cap \{p_{2s}\leq p_{1s}\}=\{p_{2s}\leq p_{1s}\} \cap \{p_{2s}\leq -w p_{1s}+w \} \cap \{0 \leq p_{1s}\leq1\} \cap \{0 \leq p_{2s}\leq1\}$ We will argue that this polytope is forward flow invariant, \textit{i.e.}, if we start from an initial condition $\vec{x} \in P'_{Hare}$
$\psi (t, \vec{x}) \in P'_{Hare} $ for all $t>0$. On the $p_{1s}, p_{2s}$ subspace $P'_{Hare}$ defines a triangle with vertices $A=(0, 0)$, $B=(1, 0)$ and $C=(\frac{w}{w+1}, \frac{w}{w+1})$ (see figure~\ref{fig:game13}). The line segments $AB$, $AC$ are trivially flow invariant. Hence, in order to argue that the $ABC$ triangle is forward flow invariant, it suffices to show that everywhere along the line segment $BC$ the vector field does not point ``outwards'' of the $ABC$ triangle. Specifically, we need to show
that
 for every point $p$ on the line segment $BC$ (except the Nash equilibrium $C$),
 $\frac{|\zeta_{1s}(\vec{p})|}{|\zeta_{2s}(\vec{p})|}\geq\frac{1}{w}$.

 \begin{eqnarray*}
 \frac{|\zeta_{1s}(\vec{p})|}{|\zeta_{2s}(\vec{p})|}&=& \frac{p_{1s}|p_{2s}-(p_{1s}p_{2s}+w(1-p_{1s})(1-p_{2s}))|}{p_{2s}|p_{1s}-(p_{1s}p_{2s}+w(1-p_{1s})(1-p_{2s}))|}= \frac{p_{1s}(1-p_{1s})(w-(w+1)p_{2s})}{p_{2s}(1-p_{2s})(-w+(w+1)p_{1s})}
 \end{eqnarray*}

\noindent
However, the points of the line passing through $B, C$ satisfy $p_{2s}=w(1-p_{1s})$.

 \begin{eqnarray*}
 \frac{|\zeta_{1s}(\vec{p})|}{|\zeta_{2s}(\vec{p})|}&=&  \frac{wp_{1s}(1-p_{1s})(1-(w+1)(1-p_{1s}))}{w(1-p_{1s})(1-w(1-p_{1s}))(-w+(w+1)p_{1s})}\\
  &=& \frac{p_{1s}(-w +(w+1)p_{1s})}{(1-w+ wp_{1s})(-w+(w+1)p_{1s})}\\
  &=&  \frac{p_{1s}}{1-w+ wp_{1s}}\geq  \frac{p_{1s}}{wp_{1s}} =\frac{1}{w}
 \end{eqnarray*}

We have established that the $ABC$ triangle is forward flow invariant. Since the $w$-coordination game is a potential game, all but a zero measurable set of initial conditions
converge to one of the two pure equilibria. Since $ABC$ is forward invariant, all but a zero measure of initial conditions converge to $(Hare, Hare)$.
A symmetric argument holds for the triangle $AB'C$ with $B'=(0, 1)$. The union of $ABC$ and $AB'C$ is equal to the polygon $P_{Hare}$, which implies
the first part of the lemma.
\medskip

Next, we will prove the claimed property for polytope $(P_{Stag})$. Again, due to symmetry, it suffices to prove the property for the polytope $P'_{Stag} = P_{Stag} \cap \{p_{2s}\leq p_{1s}\}=\{p_{2s}\leq p_{1s}\} \cap \{p_{2s}\geq -p_{1s} +\frac{2w}{w+1} \} \cap \{0 \leq p_{1s}\leq1\} \cap \{0 \leq p_{2s}\leq1\}$ We will argue that this polytope is forward flow invariant. On the $p_{1s}, p_{2s}$ subspace $P'_{Stag}$ defines a triangle with vertices $D=(1, \frac{w-1}{w+1})$, $E=(1, 1)$ and $C=(\frac{w}{w+1}, \frac{w}{w+1})$. The line segments $CD$, $DE$ are trivially forward flow invariant. Hence, in order to argue that the $CDE$ triangle is forward flow invariant, it suffices to show that everywhere along the line segment $CD$ the vector field does not point ``outwards'' of the $CDE$ triangle (see figure~\ref{fig:game13}) . Specifically, we need to show
that
 for every point $p$ on the line segment $CD$ (except the Nash equilibrium $C$),
 $\frac{|\zeta_{1s}(\vec{p})|}{|\zeta_{2s}(\vec{p})|}\leq 1$.

 \begin{eqnarray*}
 \frac{|\zeta_{1s}(\vec{p})|}{|\zeta_{2s}(\vec{p})|}&=& \frac{p_{1s}|p_{2s}-(p_{1s}p_{2s}+w(1-p_{1s})(1-p_{2s}))|}{p_{2s}|p_{1s}-(p_{1s}p_{2s}+w(1-p_{1s})(1-p_{2s}))|}= \frac{p_{1s}(1-p_{1s})(w-(w+1)p_{2s})}{p_{2s}(1-p_{2s})(-w+(w+1)p_{1s})}
 \end{eqnarray*}

\noindent
However, the points of the line passing through $C, D$ satisfy $p_{2s}=-p_{1s}+\frac{2w}{w+1}$.

 \begin{eqnarray*}
 \frac{|\zeta_{1s}(\vec{p})|}{|\zeta_{2s}(\vec{p})|}&=&  \frac{p_{1s}(1-p_{1s})(-w+(w+1)p_{1s})}{(-p_{1s}+\frac{2w}{w+1})(-\frac{w-1}{w+1}+p_{1s})(-w+(w+1)p_{1s})}\\
 &=&  \frac{p_{1s}(1-p_{1s})}{(-p_{1s}+\frac{2w}{w+1})(-\frac{w-1}{w+1}+p_{1s})}= \frac{p_{1s}(1-p_{1s})}{\frac{2(w-1)}{w+1}(-\frac{w}{w+1}+p_{1s})+p_{1s}(1-p_{1s})} \leq 1
 \end{eqnarray*}

 We have established that the $CDE$ triangle is forward flow invariant. Since the $w$-coordination is a potential game, all but a zero measurable set of initial conditions
converge to one of the two pure equilibria. Since $CDE$ is forward invariant, all but a zero measure of initial conditions converge to $(Stag, Stag)$.
A symmetric argument holds for the triangle $CD'E$ with $D'=(\frac{w-1}{w+1}, 1)$. The union of $CDE$ and $CD'E$ is equal to the polygon $P_{Stag}$, which implies
the second part of the lemma.
\end{proof}

\begin{proof}
The measure/size of $\mu(P_{Hare})= 2 |ABC|= \frac{w}{w+1}$, and similarly the measure of   $\mu(P_{Stag})= 2 |CDE|= \frac{2}{(w+1)^2}$.
The average limit performance of the replicator satisfies $\int_{g(\Delta)} sw(\psi(x)) d \mu\geq 2w\cdot \mu(P_{Hare})+2 \big(1-\mu(P_{Hare})\big)=2\frac{w^2+1}{w+1}$.
Furthermore,  $\int_{g(\Delta)} sw(\psi(x)) d\mu\leq 2w\big(1- \mu(P_{Stag})\big)+2\cdot \mu(P_{Stag})= 2w (1-\frac{2}{(w+1)^2}) +2\cdot \frac{2}{(w+1)^2}=2w - 4\frac{w-1}{(w+1)^2}$.
This implies that $\frac{w(w+1)^2}{w(w+1)^2-2w+2} \leq APoA \leq \frac{w^2+w}{w^2+1}$.
\end{proof}

\subsection{Analysis of N-star graph}
\label{appendix:Star}

{\bf Notation.}
To simplify notation in  this section, we rename strategy $Stag$ as strategy $A$ and strategy $Hare$ as strategy $B$.
Let's consider a mixed strategy profile as $(\bf{x_1},\dots, \bf{x_n}, \bf{y})$ where $\bf{x_i}$ denotes the mixed strategy of ``leaf" agent $i$, and $\bf{y}$ denotes the mixed strategy of the center agent. Since it suffices  to track for each agent the probability with which they play strategy $A$, \textit{i.e.}, \textit{Hare}, we will sometimes abuse  notation and denote the mixed strategy of agent $i$ by $x_i\defeq x_{iA}$, i.e., the probability with which he is playing strategy $A$ as well as denote a mixed strategy profile by $(x_1,\dots, x_n, y)$.

\medskip
``leaf"
Here is the high level idea of the analysis: We start  by showing that the only fixed points with region of attraction with positive measure are those in which all agents choose strategy $A$ or all agents choose strategy $B$. After that we show that the limit point of any nontrivial trajectory will be either one of the two mentioned, or a %fully
mixed Nash.
Therefore we need to compute the regions of attraction of the two fixed points where all choose $A$ or all choose $B$. To do that, we need to compute the boundary of these two regions (namely the Center-Stable manifold of the fully mixed ones). This happens as follows: Given an initial point $(x_{1}(0),\dots,x_{n}(0),y(0))$, we compute the possible fully mixed limit point $(x_{1},\dots,x_{n},\frac{w}{w+1})$ (there can only be one such possible limit point due to point-wise convergence which is furthermore constrained  due to Lemma \ref{inv-star} below) that is on the boundary of the two regions. If the initial condition is on the upper half space w.r.t to the possible fully mixed limit point $(x_{1},\dots,x_{n},\frac{w}{w+1})$ the dynamics converge to the everyone playing $A$, otherwise to everyone playing $B$.

\subsubsection{Structure of fixed points}\label{structure}

If a ``leaf"  agent $i$ applies a randomized/mixed strategy at a fixed point, it must be the case that the strategy of the center agent $y = \frac{w}{w+1}$. Otherwise, the ``leaf" agent would strictly prefer either strategy $A$ or strategy $B$.
% because $u_{iA} = u_{iB}$ and $u_{iA} = y$ and $u_{iB} = w(1-y)$ ($u_{iA},u_{iB}$ are the expected utilities for agent $i$ given that he chooses $A,B$ respectively with probability 1).
Hence the fixed points of the star graph game have the following structure: If the center agent has a pure strategy, then all agents must be pure. If the center agent has a mixed strategy, then $\sum_{i}x_{i} = \frac{w}{w+1}n$. In that case, if all the ``leaf"  agents have pure strategies then $y$ can have any value in $[0,1]$, otherwise $y = \frac{w}{w+1}$.

\subsubsection{Invariants}
\begin{lemma}\label{inv-star} $[\ln(x_{i}(t))-\ln(1-x_{i}(t))]-[\ln(x_{j}(t))-\ln(1-x_{j}(t))]$ is invariant for all $i,j$ (independent of $t$).
\end{lemma}
\begin{proof}
$\frac{d}{dt}[\ln(x_{i}(t))-\ln(1-x_{i}(t))]-\frac{d}{dt}[\ln(x_{j}(t))-\ln(1-x_{j}(t))] = [y-w(1-y)]-[y-w(1-y)]=0.$
\end{proof}

Next, we will argue that if we start in the interior of $\Delta$, the system can converge to fixed points,  where either all agent play A or B, or to a fully mixed Nash where $y = \frac{w}{w+1}$ and $\sum x_{i} = \frac{w}{w+1}n$.

\begin{lemma}\label{ABmixed}
For all initial conditions in the interior of $\Delta$, either the dynamic converges to all A's, i.e, (1,\dots,1), or to all B's, i.e., (0,\dots, 0), or to some fully mixed fixed point, i.e, $(x_1,\dots, x_n, \frac{w}{w+1})$ with $0<x_i<1$ for all $i$, and $\sum_i x_{i} = \frac{w}{w+1}n$.
\end{lemma}
\begin{proof}
We consider the following two cases:
\begin{itemize}
\item If $x_{i}(t) \to 1$ for some $i$, then $\ln(x_{i}(t))-\ln(1-x_{i}(t)) \to +\infty$. So from Lemma \ref{inv-star} for every $j$ we get that $\ln(x_{j}(t))-\ln(1-x_{j}(t)) \to +\infty$, hence $x_{j}(t) \to 1$. Due the structure of the equilibrium set and pointwise convergence, $y(t)$ must converge to $0$ or $1$. Due the fact that the fixed point $(1,\dots,1,0)$ is repelling we get that the system converges to all A's. The same argument is used if $x_{i}(t) \to 0$ for some $i$.
\item If the dynamic converges to an equilibrium were all ``leaf" agents are mixed, then $y = \frac{w}{w+1}$ and  $\sum_i x_{i} = \frac{w}{w+1}n$ because by the analysis of the structure of the fixed points that is the only possibility.
\end{itemize}
\end{proof}

Let $(x_{1}(0),\dots,x_{n}(0),y(0))$ be the initial condition, where $x_{i}(0),y(0)$ are the probabilities agent $i$, center agent choose $A$ ($1-x_{i}(0),1-y(0)$ will be the probability to choose $B$) respectively. By Lemma \ref{ABmixed}, we know that the corresponding trajectory will converge either to the all A's equilibrium or the all B's equilibrium or a fully mixed one. Next, by using Lemma \ref{inv-star} we will narrow down the possibilities for this fully mixed equilibrium to a single one, which we denote by $(x_{1},\dots,x_{n},\frac{w}{w+1})$.
 %that the trajectory can converge to apart from the all A's or all B's.

 For each leaf agent $i>1$, we define a positive constant $c_{i}$ such that $$c_{i} =\frac{x_{i}(0)/(1-x_{i}(0))}{x_{1}(0)/(1-x_{1}(0))}.$$

 Due to Lemma \ref{inv-star} the quantity $\frac{x_{i}(t)/(1-x_{i}(t))}{x_{1}(t)/(1-x_{1}(t))}$ is time invariant.
 Hence, the limit point $(x_{1},\dots,x_{n},\frac{w}{w+1})$ must satisfy this condition, i.e,

 \begin{align}\label{system}
x_{i} &= \frac{c_{i}x_{1}}{1+(c_{i}-1)x_{1}} %\\\label{system2}
%x_{1}'&\left(\sum\frac{c_{i}}{1+(c_{i}-1)x_{1}'}\right) = \frac{w}{w+1}n
\end{align}

 \noindent
Moreover, by Lemma \ref{ABmixed} it must satisfy $\sum x_{i} = \frac{w}{w+1}n$, which combined with (\ref{system}) implies that:

\begin{align}
%x_{i}' &= \frac{c_{i}x_{1}'}{1+(c_{i}-1)x_{1}'} \\
\label{system2}
\sum_i\frac{c_{i}x_{1}}{1+(c_{i}-1)x_{1}} = \frac{w}{w+1}n
\end{align}

\noindent
where we have defined $c_1=1$.

Observe that the function $f(x) = \frac{cx}{1+(c-1)x}$ is strictly increasing in $[0,1]$ (given any fixed positive $c$)  and $f(0)=0,f(1)=1$. Therefore $g(x) = \sum_i\frac{c_{i}x}{1+(c_{i}-1)x} - \frac{w}{w+1}n$ is strictly increasing in $[0,1]$ (as sum of strictly increasing functions in $[0,1]$) and $g(0)=-\frac{w}{w+1}n<0$ and $g(1) = n - \frac{w}{w+1}n>0$. Thus, it has always a unique solution in $[0,1]$ and equivalently the system of equations (\ref{system},\ref{system2}) has a unique solution. Together with $y=\frac{w}{w+1}$, the equilibrium limit point lies in the interior of $\Delta$. Given $x_{1}(0),\dots,x_{n}(0)$ we can compute (approximate with arbitrary small error $\epsilon$) $x_{1},\dots,x_{n}$ via binary search (using Bolzano's theorem).
\begin{lemma}\label{kl-star} Since star graph is a bipartite graph from Lemma \ref{lemma:KL} we have that since $(x_1,\dots,x_n,y)$ is a fully mixed Nash then along any system trajectory $((x_1(t),\dots,x_n(t),y(t)))$ the function $$\frac{w}{w+1}\ln(y(t))+\frac{1}{w+1}\ln(1-y(t))-\sum_{i}[ x_{i}\ln(x_{i}(t))+(1-x_{i})\ln(1-x_{i}(t))]$$ is (time) invariant, i.e. independent of $t$.
\end{lemma}
\begin{lemma} If $y(t)\geq (>)\frac{w}{w+1}$ and $\sum x_{i}(t)> (\geq)\frac{w}{w+1}n$ for some $t$, the trajectory converges to all $A$'s and if $y(t)\leq (<)\frac{w}{w+1}$ and $\sum x_{i}(t)< (\leq)\frac{w}{w+1}n$ for some $t$, the trajectory converges to all $B$'s.
\end{lemma}
\begin{proof}
%This is quite easy.
 In the first case, $y(t)$ is increasing and $x_{i}(t)$ (for all $i$) are non-decreasing and thus $y(t')>\frac{w}{w+1}$ and $\sum x_{i}(t')>\frac{w}{w+1}n$  holds for all $t' > t$. In the second case $y(t)$ is decreasing and $x_{i}(t)$ (for all $i$) are non-increasing and thus $y(t')<\frac{w}{w+1}$ and $\sum x_{i}(t')<\frac{w}{w+1}n$ holds for all for $t' > t$. Combining this with Lemma \ref{ABmixed}, concludes the proof when we consider the first combination of inequalities. The second combination of inequalities follows in a similar manner.
\end{proof}
Therefore if a trajectory converges to  the fully mixed equilibrium $(x_{1},\dots,x_{n},\frac{w}{w+1})$ then at any time $t$ we must have $\sum x_{i}(t)>\frac{w}{w+1}n$ and $y(t)<\frac{w}{w+1}$ ($x_{1}(t),\dots,x_{n}(t)$ are decreasing and $y(t)$ increasing) or $\sum x_{i}(t)<\frac{w}{w+1}n$ and $y(t)>\frac{w}{w+1}$ ($x_{1}(t),\dots,x_{n}(t)$ are increasing and $y(t)$ decreasing).
\begin{comment}
Finally looking at the Jacobian of the matrix for the fixed point $(x_{1}',\dots,x_{n}',\frac{w}{w+1})$, it follows that it has exactly one negative eigenvalue, namely we have one dimensional stable manifold.
\end{comment}
Combining all the facts together, we get that the stable manifold of the fixed point $(x_{1},\dots,x_{n},\frac{w}{w+1})$ can be described as follows: $(x_{1}(0),\dots,x_{n}(0),y(0))$ lies on the stable manifold if $\sum_{i}x_{i}(0)>n\frac{w}{w+1}$ and $y(0)<\frac{w}{w+1}$ or $\sum_{i}x_{i}(0)<n\frac{w}{w+1}$ and $y(0)>\frac{w}{w+1}$ and by Lemma \ref{kl-star} we get that \begin{equation}\label{eq:y0}
y(0)^{\frac{w}{w+1}}(1-y(0))^{\frac{1}{w+1}}=c \prod_{i} x_{i}(0)^{x_{i}}(1-x_{i}(0))^{1-x_{i}},
\end{equation}
where $c = \frac{\left(\frac{w}{w+1}\right)^{\frac{w}{w+1}}\left(\frac{1}{w+1}\right)^{\frac{1}{w+1}}}{\prod_{i} (x_{i})^{x_{i}}(1-x_{i})^{1-x_{i}}}$.
\begin{lemma}\label{root-star} The function $x^w(1-x) $ is strictly increasing in $[0,\frac{w}{w+1}]$ and decreasing in $[\frac{w}{w+1},1]$.
\end{lemma}

\noindent
By Lemma \ref{root-star} we have that there exist at most two $y(0)$ that satisfy (\ref{eq:y0}), one which is $ \geq \frac{w}{w+1}$ and one that $\leq \frac{w}{w+1}$. If $\sum x_{i}(0) < n\frac{w}{w+1}$, $y(0)$ should be the largest root of the two so that dynamics converges to the fully mixed, otherwise the smallest root. If now the initial condition $y(0)$ does not satisfy (\ref{eq:y0}), then the dynamics converges to all $A$'s if $y(0)$ is greater that it is supposed (so that dynamics converges to the fully mixed) and to all $B$'s otherwise. Therefore we have the oracle below:

\subsubsection{Oracle Algorithm}
\begin{center}
\fbox{\parbox{5.5in}{
\begin{minipage}{5.4in}
\begin{tt}
{\bf Oracle}
\begin{enumerate}
\item Input: Initial condition $(x_{1}(0),\dots,x_{n}(0),y(0))$
\item Output: A or B or mixed
\item If $\sum x_{i}(0)>(\geq)\frac{w}{w+1}n$ and $y(0)\geq (>)\frac{w}{w+1}$ return A.
\item If $\sum x_{i}(0)<(\leq)\frac{w}{w+1}n$ and $y(0)\leq (<)\frac{w}{w+1}$ return B.
\item Compute by solving system \ref{system}-\ref{system2} $x_{1},\dots,x_{n}$ (binary search)
\item Let $f(\rho) = \left(\frac{\rho(w+1)}{w}\right)^{\frac{w}{w+1}}\left[(1-\rho)(w+1)\right]^{\frac{1}{w+1}}- \prod_{i} \left(\frac{x_{i}(0)}{x_{i}}\right)^{x_{i}}\left(\frac{1-x_{i}(0)}{1-x_{i}}\right)^{1-x_{i}}$
\item if ($\sum_{i}x_{i}(0)>\frac{w}{w+1}n$ and $f(y(0))<0$) or \\($\sum_{i}x_{i}(0)<\frac{w}{w+1}n$ and $f(y(0))>0$) return B.
\item if ($\sum_{i}x_{i}(0)>\frac{w}{w+1}n$ and $f(y(0))>0$) or \\($\sum_{i}x_{i}(0)<\frac{w}{w+1}n$ and $f(y(0))<0$) return A.
\item return mixed fixed point $(x_1,\dots,x_n,\frac{w}{w+1})$
\end{enumerate}
\end{tt}
\end{minipage}
}}
\end{center}
\textbf{Remark:} Given any point from $\Delta$ uniformly at random, under the assumption of solving exactly the equations to compute $x_{1},\dots,x_{n}$ to arbitrary high precision the probability that the oracle above returns mixed is zero.

\subsection{Proof of Corollary \ref{thm:1868}}

\begin{proof} We first prove the following claim.
\begin{claim}\label{cl:overflow} $\mathbb{P}\left[\sum x_{i}(0) \leq n\frac{w}{w+1}\right]$ is increasing with $n$.
\end{claim}
\begin{proof}
We set by $X_1,...,X_n$ the random starting points $x_i(0),...,x_n(0)$. Let $S_n:=\sum_{i=1}^n X_i$, $x\in\mathbb R$, $n=2,3,\dots$, and
\begin{equation*}
	G_n(x):=P(S_n/n\le x)=\frac1{n!}\,\sum_j(-1)^j\binom nj (nx-j)_+^n,
\end{equation*}
where $u_+:= \max (0,u)$; (Irwin--Hall distribution). We may assume that the summation above is over all integers $j$, where $\binom nj=0$ if $j\notin\{0,\dots,n\}$. Let
\begin{equation} 
	D_n(x):=G_{n+1}(x)-G_n(x).
\end{equation}
It suffices to show that $D_n(\frac{w}{w+1})\ge 0$ for $w\ge 1$.

First observe that $D_n$ is $n-1$ times continuously differentiable, $D_n=0$ outside $(0,1)$, and also holds that $D_n(1-x)=-D_n(x)$ (by symmetry).
For small $x>0$ we get $G_n(x)=\frac1{n!}\,n^nx^n>\frac1{(n+1)!}\,(n+1)^{n+1}x^{n+1}=G_{n+1}(x)$, whence $D_n(x)<0$, whence $D_n>0$ in a left neighborhood of $1$, with $D_n(1)=0$ and $D_n(1/2)=0$. It suffices to show that $D_n$ has no roots in $(1/2,1)$ since $1>\frac{w}{w+1} \geq 1/2$.

Suppose the contrary. By the symmetry, $D_n$ has at least $3$ roots in $(0,1)$. Then, by Rolle theorem, the derivative $G_n^{(n-1)}$ of $G_n$ of order $n-1$ has at least $3+n-1=n+2$ roots in $(0,1)$. This contradicts the fact (we prove below) that, for each integer $j$ such that $n-1\ge j\ge\frac{n-1}2$, $D_n$ has exactly one root in interval
$$h_{n,j}:=[\tfrac jn,\tfrac{j+1}n).$$
\noindent
Indeed, take any $x\in[1/2,1]$. It is not hard to see that, for $j_x:=j_{n,x}:=\lfloor nx\rfloor$,
\begin{equation*}
	G_n^{(n-1)}(x)=n^{n-1}\sum_{j=0}^{j_x}(-1)^j\binom nj (nx-j)	
\end{equation*}		
\begin{equation*}
	=\frac{n^{n-1}}{n-1}(-1)^{j_x}\binom n{j_x+1}({j_x}+1) ((n-1)x-{j_x}).
\end{equation*}
Similarly, for $k_x:=j_{n+1,x}$,
\begin{equation*}
	G_{n+1}^{(n-1)}(x)=\frac{(n+1)^{n-1}}{2n(n-1)}
(-1)^{k_x}\binom{n+1}{k_x+1}(k_x+1)P(n,k_x,x),
\end{equation*}
where
\begin{equation*}
	P(n,k,x):=-2 k \left(n^2-1\right) x+k (k n-1)+n \left(n^2-1\right) x^2.
\end{equation*}
Note that $k_x\in\{j_x,j_x+1\}$. We consider the following two cases.

\bigskip
\noindent
Case 1: $k_x=j_x=j\in[\frac{n-1}2,n-1]$, which is equivalent to $x\in h'_{n,j}:=[\frac jn,\frac{j+1}{n+1})$. It also follows that in this case $j\ge n/2$. In this case, it is not hard to check that $D_n(x)$ equals $(-1)^j P_{n,j,1}(x)$ in sign, where
\begin{align*}
	P_{n,j,1}(x):=& 2 j n^n (n-j)+x \left(-2 (n-1) n^n (n-j)-2 j (n+1)^n \left(n^2-1\right)\right)+ \\& j (n+1)^n (j n-1)+(n+1)^n
   \left(n^2-1\right) n x^2,
\end{align*}
which is convex in $x$. Moreover, $P_{n,j,1}(\frac jn)$ and $P_{n,j,1}(\frac{j+1}{n+1})$ each equals $2 n^n - (1 + n)^n<0$ in sign. So, $P_{n,j,1}<0$ and hence $D_n$ has no roots in $h''_{n,j}=[\frac jn,\frac{j+1}{n+1})$.

\bigskip
\noindent
Case 2: $k_x=j+1$ and $j_x=j\in[\frac{n-1}2,n-1]$, which is equivalent to $x\in h''_{n,j}:=[\frac{j+1}{n+1},\frac{j+1}n)$. In this case, it is true that $D_n(x)$ equals $(-1)^{j+1} P_{n,j,2}(x)$ in sign, where
\begin{align*}
	P_{n,j,2}(x):=& (j+1) \left((n+1)^n (j n+n-1)-2 j n^n\right)+2 (j+1) \left((n-1) n^n-(n+1)^n \left(n^2-1\right)\right) x+ \\& n
   \left(n^2-1\right) (n+1)^n x^2,
\end{align*}
and is convex in $x$. Moreover, $P_{n,j,1}(\frac{j+1}n)$ and $-P_{n,j,1}(\frac{j+1}{n+1})$ equal $2 n^n - (1 + n)^n<0$ in sign. So, $P_{n,j,2}$ has exactly one root in $h''_{n,j}$ and hence so does $D_n$.

Since the interval $h_{n,j}$ is the disjoint union of $h'_{n,j}$ and $h''_{n,j}$, $D_n$ has exactly one root in $h_{n,j}=[\tfrac jn,\tfrac{j+1}n)$, and the proof is complete.
\end{proof}
There are exactly two possible outcomes with positive probability; all the agents choose strategy $A$ and all choose strategy $B$. Assume we take one sample at random $(x_1,\dots,x_n,y)$ from $\times_{i=1}^{n+1}\Delta_2$ where $n+1$ are the number of agents. Let $p_x, p_y$ be the the probability that a sample at random satisfies $\sum_i x_i \leq n\frac{w}{w+1}$ and $y< \frac{w}{w+1}$ respectively.  It turns out from the oracle above on the star-graph game (see also discussion earlier) that if ($\sum_i x_i < n\frac{w}{w+1}$ and $y< \frac{w}{w+1}$) or ($\sum_i x_i < n\frac{w}{w+1}$ and $f(y)>0$) or ($\sum_i x_i > n\frac{w}{w+1}$ and $f(y)<0$) then the dynamics eventually converges to all agents choose $B$. Hence the region of attraction of the outcome all agents choose $B$ will be at least the probability $p_x \cdot p_y \geq \frac{w}{w+1} \cdot \frac{w}{w+1}$ (by Claim \ref{cl:overflow} for $n=1$ is minimum). Since the optimal is $w(n+1)$, we get that the average price of anarchy  is at most $\frac{w(n+1)}{(\frac{w}{w+1})^2 w(n+1) + (1-(\frac{w}{w+1})^2)(n+1)} = \frac{w(w+1)^2}{w^3+2w+1}$. The quantity $\frac{w(w+1)^2}{w^3+2w+1}$ is less than $1.42$.
\end{proof}

%\section{Average case analysis of linear symmetric load balancing games}
\subsection{Proof of Theorem \ref{thm:AvePoA}}

In this section, we will prove the following bounds on the average price of anarchy of linear, symmetric load balancing games.
We will break down the proof of theorem  \ref{thm:AvePoA} into several technical lemmas.
The next definition encodes Nash equilibria where randomizing agents do not ``interact'' with each other.

\begin{definition}
We call a mixed Nash equilibrium of a load balancing game to be almost pure, if the intersection of the supports of the strategies of any two randomizing agents contains  only edges whose latency functions are constant functions. %empty.
\end{definition}

\begin{lemma}
The average price of anarchy of a symmetric, linear load balancing games is at most equal to the ratio of the cost of the worst almost pure Nash equilibrium divided by the cost of the optimal outcome.
\end{lemma}

\begin{proof}
By corollary \ref{col:weakly} we have that for all but a zero measure of initial conditions replicator dynamics converges to weakly stable equilibria. By definition, weakly stable equilibria have the property that given any two agents with mixed strategies strategies if one agent deviates to one the strategies in his  support and plays it with probability one then the second agent should still stay indifferent between the strategies in his support. If there exists two agents with  mixed strategies such that the intersection of their supports contains machines with strictly increasing latency functions then if one agent deviates to playing that machine with probability one, he will strictly increase the cost experienced by the second agent on that machine, whereas by this deviation he can only decrease the cost of all other machines in the support of the second agent. The second agent is no longer indifferent between the strategies in his support and thus the initial equilibrium was not weakly stable. In the worst case average price of anarchy places all of the probability mass of initial conditions to the worst almost pure Nash equilibrium. In this case the average price of anarchy would be equal to the ratio of the cost of the worst almost pure Nash equilibrium divided by the cost of the optimal outcome.
\end{proof}

\begin{lemma}
\label{lem:opt_pure}
In symmetric linear load balancing games all pure Nash equilibria have optimal makespan.
\end{lemma}

\begin{proof}
Suppose not, that is, suppose that there exists a pure Nash equilibrium whose makespan, i.e., the load of the most congested machine, is not optimal amongst all outcomes/configurations. That means that its most loaded machine must be a machine with a strictly increasing cost function that has higher load than its load at the optimal outcome.\footnote{If there exist more than one outcomes with minimal makespan, we just arbitrary focus on one of the optimal configurations.} Hence, there must be another machine whose load is strictly less than its load at the optimal configuration. If we move one agent from the first to the second machine we claim that its cost will strictly decrease. Indeed, its new latency is at most the latency of the second machine in the optimal configuration, which is less or equal to the optimal makespan, which by hypothesis is strictly less than the makespan of the first configuration, which was its original cost. Hence, the original configuration cannot be a Nash equilibrium and we have reached a contradiction.
\end{proof}

\begin{lemma}
In any symmetric, linear load balancing games the ratio of the cost of the worst almost pure Nash equilibrium divided by the cost of the optimal outcome is at most 3/2. Furthermore, this bound is tight.
\end{lemma}

\begin{proof}
First, we create the lower bound. We have a load balancing game with two agents and three machines. The latency function for the first machine is $3x$ whereas for the other two machines is $2x$. It is straightforward to check that the strategy outcome where the first agent chooses the first machine and the second agent chooses one of the remaining two machines uniformly at random is a Nash equilibrium and, in fact, a  weakly stable one. The makespan of this equilibrium is $3$, whereas the optimal state has each of the two agents choosing deterministically one of the last two machines and using it by themselves. The makespan of that outcome is  $2$, which results in a lower bound of $3/2$.\footnote{This construction is due to Bobby Kleinberg.}

Next, we will show that this bound is tight. First, we will establish that it suffices to examine Nash equilibria where the intersection between the supports of the mixed strategies of any two randomizing agents is empty. Indeed, suppose that we have two randomizing agents where the intersection of their supports contains some machines with constant latency functions. If we force one of the two agents to deviate and choose deterministically the strategy of constant latency in his support then the makespan of the state remains constant and furthermore the outcome is still a weakly stable Nash. The reason that it remains a Nash is that if an agent wished to deviate to some strategy used by the deviating agent originally, then when deviating to that machine he would experience exactly the same cost as when using the machine with the constant cost function. Thus, he could have profitably deviated in the initial configuration. This is impossible since that configuration was a Nash equilibrium. Trivially, this new Nash equilibrium is weakly stable since we  have only decreased the number of randomizing agents and the supports of the remaining randomizing agents remained the same. We can keep performing these deviations up until there no longer randomizing agents for which the intersection of the supports contains any machine (of constant latency function). Hence, in terms of identifying the almost pure Nash equilibrium with the worst makespan it suffices to focus on the set of almost mixed Nash equilibria where the intersection of the supports of any two randomizing agents is empty.

We have established that if suffices to focus on mixed Nash equilibria where each machine has at most one randomizing agent.
We will establish that the makespan of each such equilibrium is within a $3/2$ factor  of the makespan of a pure Nash equilibrum, which by Lemma \ref{lem:opt_pure} implies that it is within a $3/2$ factor of the optimal makespan. The argument is as follows: We will start from the mixed Nash and will proceed by fixing the randomizing agents to playing strategies  in their support with probability one. We start
from the randomizing agent $i$ that experiences minimum cost amongst all randomizing agents.
We fix him to playing the strategy in his support that he chose with minimal probability in the original mixed Nash.
We also fix the rest of the randomizing agents to arbitrary strategies in their support. Next, we repeatedly go through all agents in decreasing cost order and we allow each agent to move and migrate to the least expensive (available) machine if it is strictly cheaper than his current machine. Due to symmetry once we find one agent who does not wish deviate all of the rest of the agents do not wish to deviate either due to symmetry of the available machines.
This process will terminate at equilibrium since this is a potential game. Furthermore, agent $i$ (nor of any of the other agents in his machine) will ever move during this process. If he did move then there would exist at some point a profitable deviating move from him. However, immediately after fixing the randomizing agents to choosing something in their current support, agent $i$ did not have any improving deviations since his experienced cost was minimal amongst all randomizing agents and hence at least as small as the cost of any deviation.
In fact, the cheapest available deviations are exactly the strategies that belonged in his support. As we allow costly agents to move greedily from their current strategy to the best available strategy the cost of the best available deviation cannot decrease with time. Thus, agent $i$ will not deviate. Hence, the makespan at the resulting pure Nash equilibrium will be at least equal the cost of agent $i$ when his was fixed to the strategy that he played with minimal probability. If we denote that edge as $e$ and its load (excluding agent $i$) as $x_e$ then this implies that the makespan at the resulting Nash and thus the optimal makespan is at least $a_e(x_e+1)+b_e$. However, the original mixed state was an equilibrium and if agent $i$ played strategy $e$ with probability $p$ then no agent in the original Nash equilibrum would experience cost more than $a_e(x_e+p+1)+b_e$.\footnote{If he did we would strictly prefer to deviate to edge $e$.} But since $e$ was chosen to be the strategy played with minimal probability in his original support $p\leq 1/2$ and hence no agent can experience cost more than  $a_e(x_e+3/2)+b_e$. So, the original makespan is at most  $a_e(x_e+3/2)+b_e$ and the optimal makespan is at least  $a_e(x_e+1)+b_e$. The ratio between these two terms becomes maximal (and equal to $3/2$)  for $b_e=0$ and $x_e=0$, which is exactly satisfied by our tight lower bound.
\end{proof}

\begin{remark} If we slightly perturb the above tight example so that  the latency function for the first machine is $3x$ whereas for the other two machines is $2x+\epsilon$ then the continuum of equilibria with the bad makespan will have a non-negligible region of attraction resulting in an average price of anarchy which is strictly greater than one.
\end{remark}
%\begin{comment}
%\begin{proof}
%\end{proof}

\begin{lemma}
In generic symmetric linear load balancing games  the set of almost pure Nash equilibria coincides with the set of pure Nash equilibria.
Specifically, %given any number of agents and machines,
 the set of linear latency functions such the set of almost pure Nash equilibria is a strict superset of the set of pure Nash equilibria is of measure zero within the set of all linear latency functions.
\end{lemma}

\begin{proof}
We will show that if a linear symmetric load balancing game has an almost pure Nash equilibrium that is not pure, i.e., that has at least one agent using a randomized strategy, then the coefficients of the linear latency functions belong to  a zero measure set. Indeed, let's focus on one of the randomizing agents. Since this agent is indifferent between (at least) two machines/edges $e,e'$ and he is the only randomizing agent using these machines (or some of these machines have a constant latency function) then there exist integer numbers $k,k'$, so that the cost of these two machines are equal under loads $k,k'$. This implies that $a_e\cdot k+b_e=a_{e'}\cdot k'+b_e'$. However, for any fixed $k,k'$ the set of coefficients  $a_e,a_{e'},b_e,b_{e'}$ that satisfy  this linear equation is a zero-measure set. Hence, given any number of agents and machines the set of latency functions that have almost pure Nash equilibria that are not pure can be expressed as a countable union of zero-measure sets, which is a zero-measure set.
\end{proof}

\noindent
By combining the lemmas of this section, Theorem \ref{thm:AvePoA} follows immediately.

\subsection{Proof of Theorem \ref{thm:ballsapoa}}
\label{sec:ballsbinss}
In the classic game of $N$ identical balls/agents, with $N$ identical bins/machines,
each ball chooses a distribution over the bins selfishly and we assume that the cost of bin $\gamma$ is equal to $\gamma$'s load. We know for this game that the PoA is $\Omega( \frac{\log N}{\log \log N})$ \cite{vocking}.
 We will prove that the Average PoA is $1$. This is derived via corollary \ref{col:weakly} and by showing that in this case the set of weakly stable Nash equilibria coincides with the set of pure equilibria.

\begin{lemma}\label{lem:ballsbins1} In the problem of $N$ identical balls and $N$ identical bins every weakly stable Nash  equilibrium is pure.
\end{lemma}
\begin{proof} Assume we have a weakly Nash equilibrium $\vec{p}$. From corollary \ref{col:weakly}, we have the following facts:
\begin{itemize}
\item Fact 1: For every bin $\gamma$, if a agent $i$ chooses $\gamma$ with probability $1>p_{i\gamma}>0$, he must be the only agent that chooses that bin with nonzero probability.
 Let $i,j$ two agents that choose bin $\gamma$ with nonzero probabilities and also $p_{i\gamma},p_{j\gamma}<1$. Clearly if agent $i$ changes his strategy and chooses bin $\gamma$ with probability one, then agent $j$ doesn't stay indifferent (his cost $c_{i \gamma}$ increases).
\item Fact 2: If agent $i$ chooses bin $\gamma$ with probability one, then he is the only agent that chooses bin $\gamma$ with nonzero probability. This is true because every ball/agent $j\neq i$ can find a bin with load less than 1 to choose.
\end{itemize}
From Facts 1,2 and since the number of balls is equal to the number of bins we get that $\vec{p}$ must be pure.
\end{proof}
\begin{proof}
Hence from Lemma ~\ref{lem:ballsbins1} and \ref{col:weakly} we get that for all but measure zero starting points of $g(\Delta)$, the replicator converges to pure Nash equilibria. Every pure Nash equilibrium (each ball chooses a distinct bin) has social cost (makespan) $1$, which is the optimal makespan. Hence the Average PoA is $1$.
\end{proof}

\textbf{Remark:}
The lemma below shows how crucial is Lindel\H{o}f's lemma (essentially separability of $R^m$ for all $m$) in the proof of Theorem \ref{zero}. Even simple and well studied instances of games with constant number of agents and strategies may have uncountably many equilibria. In such games, naive union bound arguments do not suffice since we cannot argue about the measure of an uncountable union of measure zero sets.

\begin{lemma} ~\label{lem:infinite1} Let $n \geq 4$ then the set of Nash equilibria of the $N$ balls $N$ bins game is uncountable.
\end{lemma}
\begin{proof}
We will prove it for $N=4$ and then the generalization is easy, i.e., if $N>4$ then the first 4 agents will play as shown below in the first 4 bins and each of the remaining $N-4$ agent will choose a distinct remaining bin. Below we give matrix $A$ where $A_{i\gamma} = p_{i\gamma}$. Observe that for any $x \in [\frac{1}{4},\frac{3}{4}]$ we have a Nash equilibrium.
\[A = \left (\begin{array}{cccc}
x & 1-x & 0 & 0\\
1/2 & 0 & 1/2 & 0\\
0& 1/2 & 0 & 1/2\\
0 & 0 & x & 1-x
\end{array}\right).\]
\end{proof}

\end{document}